\documentclass[aps,prx,superscriptaddress,twocolumn,showpacs]{revtex4-2}
\usepackage{xcolor}
\usepackage{epstopdf}
\usepackage{amsmath,amssymb}
\usepackage{commath}
\usepackage{bm}
\usepackage[colorlinks = true,
            linkcolor = blue,
            urlcolor  = blue,
            citecolor = blue,
            anchorcolor = blue]{hyperref}
            
\usepackage{cleveref}
\usepackage{graphicx}
\usepackage{xcolor}

\begin{document}
\crefname{equation}{Eq.}{Eqs.}
\crefname{figure}{Fig.}{Fig.}
\crefname{appendix}{Appendix}{Appendix}

\title{Kondo impurity in an attractive Fermi-Hubbard bath: Equilibrium and dynamics}

\author{Zhi-Yuan Wei}
\thanks{ZYW and TS contributed equally to this work.}
\address{
Max-Planck-Institut f{\"{u}}r Quantenoptik, Hans-Kopfermann-Str. 1, 85748 Garching, Germany
}%
\address{
Munich Center for Quantum Science and Technology (MCQST), Schellingstr. 4, 80799 M{\"{u}}nchen, Germany
}%

\address{Joint Quantum Institute and Joint Center for Quantum Information and Computer Science, NIST/University of Maryland, College Park, Maryland 20742, USA
}%

\author{Tao Shi\normalfont\textsuperscript{*}}
\email{tshi@itp.ac.cn}
\address{CAS Key Laboratory of Theoretical Physics and Institute of Theoretical Physics, Chinese Academy of Sciences,  Beijing 100190, China}

\author{J.~Ignacio Cirac}
\address{
Max-Planck-Institut f{\"{u}}r Quantenoptik, Hans-Kopfermann-Str. 1, 85748 Garching, Germany
}%
\address{
Munich Center for Quantum Science and Technology (MCQST), Schellingstr. 4, 80799 M{\"{u}}nchen, Germany
}%

\author{Eugene A. Demler}
\address{
Institute for Theoretical Physics, ETH Zurich, Wolfgang-Pauli-Str. 27, 8093 Zurich, Switzerland}

\date{\today}

\begin{abstract}
We investigate theoretically equilibrium and dynamical properties of a single Kondo impurity coupled to either one- or two-dimensional superconductors. We model the superconductors using the attractive Fermi-Hubbard model and employ a non-Gaussian variational approach to go beyond the conventional approximation of the constant superconducting gap. We demonstrate that dynamical properties of the system can be modified qualitatively, when space and time dependent renormalization of the superconducting gap and electron-impurity hybridization are included. In the ground state, our approach reproduces the previously known results, such as the singlet-doublet transition and $\pi$-phase shifts of the superconducting order parameter in the singlet phases, both in 1D and 2D systems. The first type of dynamics that we consider is spin dynamics following an abrupt connection of a spin-polarized impurity to a 2D superconducting electronic reservoir. We find rapid relaxation of spin polarization at the impurity site and directional emission of a magnetization pulse, which becomes damped as it propagates into the bulk. The second type of dynamics that we analyze is transport between two superconducting leads coupled through a Kondo impurity at finite bias voltage. In this setup, we go beyond analysis of the steady state to investigate full-time dynamics following an abrupt application of the bias voltage. We uncover four distinct regimes in the transient dynamics and transport properties: (I) the standard AC Josephson effect regime; (II) dynamical competition between charge-density-wave (CDW) and superconducting orders with transient Kondo correlations; (III) the coexistence of AC and DC currents facilitated by partial Kondo screening and dynamical stabilization of the superconducting order; (IV) DC Kondo transport regime modified by the superconducting order. Regime II exhibits a dynamical transition from superconducting to CDW order that locally restores the U(1) symmetry. We argue that our findings for regime IV provide a theoretical explanation for the experimentally observed anomalous enhancement of DC conductance~\cite{hata2018enhanced} and suppression of the AC Josephson current~\cite{watfa2021collapse}. Finally, we discuss how the proposed setup can be realized experimentally with ultracold atoms.

\end{abstract}

\maketitle

\section{Introduction}
\label{}
The intricate interplay between quantum magnetism and superconductivity characterizes strongly correlated electron systems. This interplay is crucial in defining the phase diagrams of various materials, including high-$T_c$ cuprates~\cite{orenstein2000advances}, heavy fermion materials~\cite{stewart1984heavy}, organic metals~\cite{organQM}, Moiré systems, and Bernal bilayer graphene~\cite{andrei2021marvels}. In these systems, superconducting and magnetic phases coexist within their phase diagrams, raising questions about whether magnetism competes with or facilitates superconductivity through non-phononic electron attraction. A notable manifestation of the superconductivity-magnetism interplay is the FFLO superconducting state~\cite{casalbuoni2004inhomogeneous}, where a spatially dependent superconducting (SC) order parameter emerges due to finite spin polarization of electrons.

In mesoscopic physics, heterostructures of ferromagnetic and superconducting materials have been studied to elucidate the interplay between these two orders and to develop new types of quantum devices~\cite{buzdin2005proximity,Franceschi2010}. Specifically, a novel quantum computing architecture has been proposed, leveraging the negative superconducting Josephson coupling across ferromagnets found in such systems~\cite{feofanov2010implementation}. A simplified model of this system reduces the magnetic region to a single quantum spin. Despite its apparent simplicity, this system exhibits a rich variety of phenomena, and it has stimulated considerable theoretical and experimental research~\cite{balatsky2006impurity,Heinrich2018}. Initial investigations into the behavior of a classical spin by Yu, Rusinov, and Shiba~\cite{yu1965bound,shiba1968classical,rusinov1969superconductivity,rusinov1969theory} revealed subgap quasiparticle states near the impurity. Subsequent studies identified a parity-changing transition~\cite{Sakurai1970,salkola1997spectral}, where the ground state alternates between an even and an odd number of electrons. Subsequent theoretical analysis demonstrated that the parity-changing transition results from the interplay between the Kondo effect and Cooper pairing~\cite{Satori1992}. When pairing is strong, the Kondo effect is suppressed by the gap in the quasiparticle spectrum, leaving the impurity spin unscreened. However, as the exchange coupling between electrons and the impurity spin strengthens, forming a Kondo singlet becomes energetically favorable, even though it involves the addition of an unpaired Bogoliubov quasiparticle. Experimentally, systems of Kondo impurities coupled to superconductors (referred to as \textit{Kondo-SC} systems) have been realized either with magnetic impurity atoms on the surface of a superconductor~\cite{yazdani1997probing,franke2011competition,hatter2015magnetic} or with quantum dots connected to superconducting electrodes~\cite{lee2017scaling}. In the latter scenario, the primary experimental probe involves measuring transport through the quantum dot~\cite{kouwenhoven1997electron}.

To date, most analyses of Kondo-SC systems rely on the following simplifications: Firstly, the superconducting order parameter has not been computed self-consistently in the presence of the impurity spin. This simplification likely fails when quasiparticle states near the impurity are significantly altered, especially when an unpaired Bogoliubov quasiparticle is added to the system. In such cases, spatially extended modifications of the SC order parameter are observed~\cite{salkola1997spectral,balatsky2006impurity,meng2015superconducting}. Secondly, theoretical analyses of problems involving quantum spins are often conducted in a 1D geometry, even when higher-dimensional systems are examined. This approach is based on arguments from standard Kondo systems (where an impurity spin is coupled to a normal Fermi liquid), where in the continuum limit, only $l=0$ fermionic states can interact with the impurity~\cite{Wilson1975,andrei1983solution,Bulla2008a,Busser2013}. However, this reduction to a 1D model is not suitable when lattice effects are relevant or when the particles in the bath interact with each other. Thirdly, the charge and magnetic fluctuations in the bath have been typically neglected, and thus the effects arising from the interplay between possible competing orders in the bath are not properly captured. Fourthly, when analyzing transport through the quantum dot, it is commonly assumed that the superconducting order parameters of the Kondo-SC system remain unaltered by non-equilibrium dynamics. This assumption holds for small applied bias voltages but is expected to break down at finite bias voltages. In addition to all that, recent experimental advances in the dynamical control of quantum dots~\cite{tsuchimoto2022large,liang2023ultrafast,burkard2023semiconductor} have enabled the observation of transient dynamics, whereas theoretical studies have been mostly limited to the analysis of steady states~\cite{avishai2003superconductor}. Even for steady states, however, certain exotic phenomena observed in recent experiments~\cite{hata2018enhanced, watfa2021collapse} remain not fully understood.

The objective of this paper is to go beyond these simplifications and analyze a Kondo impurity coupled to an electron bath when the latter feature attractive interactions. We describe the electrons using the attractive Fermi-Hubbard model, which has been employed to analyze both interacting electrons and ultracold atoms in optical lattices. In scenarios featuring an attractive Hubbard interaction, the ground state exhibits s-wave superconductivity and charge-density-wave (CDW) phenomena~\cite{scalettar1989phase,Moreo1991}, capturing the behaviour of typical s-wave superconductors and revealing a possible interplay between charge and superconducting orders. Consequently, coupling a Kondo impurity with an attractive Hubbard bath provides a new setup for probing the interplay between different types of correlations arising from the interactions between the impurity and the bath. Furthermore, owing to the experimental progress in ultracold atoms in realizing Kondo interactions~\cite{Riegger2018}, attractive Hubbard interactions~\cite{Mitra2018a,hartke2022quantum,hartke2023direct} and setups for exploring transport dynamics between superfluids~\cite{husmann2015connecting,fabritius2024irreversible,huang2023superfluid}, it is timely to initiate the study of the in- and out-of-equilibrium properties of a Kondo impurity coupled to superfluid baths. This will provide new theoretical insights needed for realizing improved hybrid superfluid-magnetic impurity devices (akin to hybrid superconductor-quantum dot devices~\cite{Franceschi2010}).

Previously, the case of a quantum spin coupled to a one-dimensional (1D) \textit{repulsive} Hubbard bath has been investigated using the density matrix renormalization group (DMRG) method~\cite{costamagna2006anderson,bragancca2021quench,cavalcante2023quench}, analyzing both ground state properties and relaxation dynamics. In contrast, existing research on the quantum spin coupled to an \textit{attractive} Hubbard bath primarily relies on Bethe ansatz methods~\cite{Zvyagin1997,Bares1999,hou1999impurity}, which calculate certain properties, such as the magnetic susceptibility, along specific integrable parameter lines. The scarcity of numerical studies on Kondo-SC systems with the attractive Hubbard bath may be attributed to the technical challenges involved in addressing this problem using standard numerical techniques. Firstly, the Hubbard interaction within the bath cannot be addressed using the standard numerical renormalization group (NRG) method~\cite{Bulla2008a}, which relies on the diagonalization of a non-interacting bath. Additionally, the spatial inhomogeneity and the strongly interacting nature of the single Kondo impurity complicate the application of dynamical mean field theory (DMFT)~\cite{georges1996dynamical} and other mean-field methods. While (time-dependent) DMRG can be employed to analyze this problem, its applicability is limited to 1D baths and short-time dynamics. As a result, the physics of the Kondo impurity coupled to the attractive Hubbard model remains largely unexplored. Developing an efficient numerical approach that can handle the Kondo singlet formation and the attractive Hubbard interaction is particularly desirable---in particular if it can also be extended beyond the 1D setting and be used to explore long-time dynamics.

\begin{figure*}
	\centering
    \includegraphics[width=1.0\textwidth]{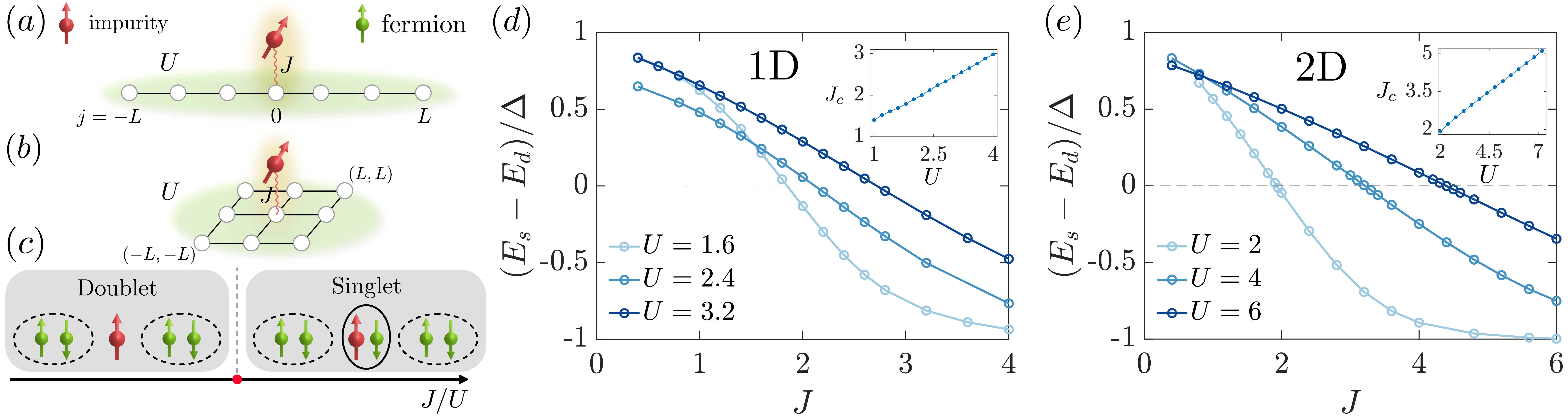}
        \caption{A Kondo impurity coupled to an attractive Fermi-Hubbard bath. (a,b) We consider a fermionic bath with an attractive Hubbard interaction of strength $U$, coupled to a Kondo impurity via a spin-exchange interaction of strength  $J$ (represented by red wavy lines). Panels (a) and (b) depict the 1D [bath size $N=2L+1$] and 2D [bath size $N=(2L+1)\times(2L+1)$] coupling geometries studied in \cref{sec_sg_12,sec_2d_relax}. These geometries are analyzed to study the ground-state properties of the system (in both 1D and 2D) and its relaxation dynamics (in 2D). (c) Singlet-doublet phase transition. When $U \gg J$, bath fermions predominantly form Cooper pairs (indicated by the dashed ovals), and the impurity behaves as a nearly free spin, resulting in a \textit{doublet} phase with a total spin $\langle \hat{S}^2 \rangle = 3/4$. Conversely, when $U \ll J$, the Kondo impurity and a bath fermion form a Kondo singlet, resulting in a \textit{singlet} phase with $\langle \hat{S}^2 \rangle = 0$. By tuning $J$ and $U$, a first-order phase transition (indicated by the red dot) occurs between the doublet and singlet phases. (d,e) The difference between the ground state energy $E_s$ in the singlet phase and $E_d$ in the doublet phase (plotted in units of the SC gap $\Delta$) as a function of the Kondo coupling $J$ for various strengths of the attractive Hubbard interaction $U$. The singlet-doublet phase transition occurs at $E_s=E_d$ (indicated by the dashed horizontal line). The results for a 1D chain ($N=401$) and a 2D square lattice ($N=41\times 41$) are shown in (d) and (e), respectively. The insets depict the critical Kondo coupling $J_c$ as a function of $U$.}
        \label{fig0}
\end{figure*}

In this work, we theoretically investigate the ground-state properties and long-time out-of-equilibrium dynamics of a Kondo impurity coupled to 1D and 2D attractive Hubbard baths [cf.~\cref{fig0,fig_mid_cur}]. Our study employs a non-Gaussian approach~\cite{Shi2018a}, which combines fermionic Gaussian states with an entangling unitary transformation between the impurity spin and the bath. We uncover a variety of phenomena in this setup, including:
\begin{enumerate}
    \item We observe the well-known singlet-doublet phase transition in the ground state~\cite{yu1965bound,shiba1968classical,rusinov1969superconductivity,rusinov1969theory,soda1967s,Satori1992,balatsky2006impurity,huang2024large}, characterized by a $\pi$ phase shift of the SC order parameter upon crossing the Kondo singlet in one dimension or at the impurity site in two dimensions. We attribute this $\pi$ phase shift to the scattering phase shift of Cooper pairs by the Kondo singlet.
    \item As an example of nonequilibrium dynamics, We analyze spin dynamics following the abrupt connection of a spin-polarized impurity to an 2D superconducting electronic reservoir. We observe rapid relaxation of the impurity polarization, accompanied by the directional emission of a pulse of spin polarization (PSP), which becomes damped as it propagates into the bulk. The directionality and the damping of the PSP are induced by both the divergent density of states and interactions among host particles.
    \item We analyze the transport dynamics between two attractively interacting leads coupled through a Kondo impurity under bias voltage, revealing a rich phase diagram with intriguing phenomena, such as the dynamical local restoration of $U(1)$ symmetry driven by the competition between charge and superconducting orders, the coexistance of AC and DC currents facilitated by partial Kondo screening~\cite{moca2021kondo} and dynamically stabilized superconduting order, and the anomalous enhancement of Kondo DC transport due to the superconducting order. Based on our analysis, we propose explicit microscopic mechanisms underlying the experimentally observed anomalous enhancement of the DC conductance~\cite{hata2018enhanced} and suppression of the AC Josephson current~\cite{watfa2021collapse}.
\end{enumerate}

This paper is organized as follows. In \cref{sec_sum}, we provide a non-technical summary of our main findings. Section \cref{sec_theo} outlines the theoretical framework of our study, including the system setup and the non-Gaussian variational approach. In \cref{sec_sg_12}, we examine the ground state properties of the system when the impurity is coupled to the central site of both 1D and 2D attractive Hubbard baths. In \cref{sec_2d_relax}, we investigate the relaxation dynamics of the impurity coupled to a 2D bath. In \cref{ld2_sec}, we systematically study the transport dynamics between two 1D leads connected by a Kondo impurity under bias voltage. Finally, we explore the potential implementation with ultracold atom platforms in \cref{sec_imple}, and present our discussion and outlook in \cref{clu}.

\section{Summary of main results}
\label{sec_sum}

Before presenting a detailed theoretical analysis, we summarize the main results of the paper at a non-technical level in this section.

\subsection{Theoretical framework}

In \cref{sec_theo}, we introduce a model of the Kondo-SC system, where a Kondo impurity is coupled to an attractive Hubbard bath. We also formulate theoretical method that will be used throughout the paper. The primary objective of this section is to extend the non-Gaussian variational approach~\cite{Shi2018a,sala2018variational,Ashida2018a,Ashida2019a} to quantum impurity models where the fermionic bath undergoes spontaneous symmetry breaking. We identify a unitary transformation that effectively decouples the impurity spin from the superconducting bath, and subsequently combine this transformation with fermionic Gaussian states to characterize both the ground state and real-time dynamics. Utilizing the time-dependent variational principle, we explore the rich quantum phases present in the ground state and non-equilibrium processes induced by intertwined orders within the system.

\subsection{Ground state properties}

In \cref{sec_sg_12}, we examine the ground state energy, spin correlations, and SC order parameters of a system where the Kondo impurity is coupled to the central site of either a 1D or 2D attractive Hubbard bath [cf.~\cref{fig0}(a,b)]. As expected~\cite{yu1965bound,shiba1968classical,rusinov1969superconductivity,rusinov1969theory,soda1967s,Satori1992,balatsky2006impurity,huang2024large}, we find that the system's ground state is governed by the competition between the Kondo coupling strength $J$ and the attractive Hubbard interaction strength $U$ [cf.~\cref{fig0}(c)]. When $J \ll U$, electrons in the bath form Cooper pairs, and the impurity spin behaves as a free spin, resulting in a doublet-phase ground state. Conversely, when $J\gg U$, the impurity forms a singlet bound state with a Bogoliubov excitation, leading to a singlet-phase ground state. Within our class of variational wavefunctions, we identify a first-order phase transition between the singlet and doublet phases as the ratio of $J$ and $U$ is varied.

In the doublet phase, the SC order only slightly decreases around the impurity, and the Kondo correlations remain weak. This indicates that the doublet phase exhibits only partial Kondo screening~\cite{moca2021kondo}. In contrast, in the singlet phase, we observe strong Kondo correlations, with the SC order significantly suppressed around the impurity. A particularly novel phenomenon occurs in the case of 1D bath, where a distinct $\pi$ phase shift of the SC order parameter in the whole bath emerges when crossing the impurity. We attribute this $\pi$ phase shift to the scattering of Cooper pairs off the Kondo singlet, where each electron acquires a $\pi/2$ phase shift due to Friedel's sum rule~\cite{langreth1966friedel,nozieres1974fermi}. In contrast, for the 2D bath, the $\pi$ phase shift of the SC order parameter occurs only around the impurity, consistently with previous observations~\cite{Flatte1997,salkola1997spectral}. This illustrates that for the interacting bath, dimensionality is important for understanding physical properties, which differs from standard quantum impurity problems, in which fermionic  baths are assumed to be non-interacting~\cite{Wilson1975,andrei1983solution,Bulla2008a,Busser2013}. Moreover, the $\pi/2$ phase shift for fermionic quasiparticles should also underlie the mechanism of the $\pi$ phase shift of the SC order parameter around the impurity in the 2D bath~\cite{balatsky2006impurity}.

\subsection{Relaxation dynamics}

In \cref{sec_2d_relax}, we investigate the long-time relaxation dynamics of a Kondo impurity in a 2D bath. Initially, the system is prepared in a product state between the impurity and the SC state of the bath fermions at half-filling. At time $t=0$, the Kondo coupling with strength $J$ is turned on. We observe the rapid relaxation of the system around the impurity toward its local equilibrium, as indicated by the growth of impurity-bath spin correlations forming a (partial) Kondo cloud. Additionally, this relaxation process induces a pulse of spin polarization (PSP) that propagates through the bath. The amplitude of this PSP is enhanced along the two diagonal directions, due to the diamond shape of the Fermi surface in the tight-binding model, with Fermi velocities pointing in the diagonal directions~\cite{gonzaleztudela17a,windt2024fermionic}. Furthermore, the SC gap in the bath leads to an exponential decay of the PSP wavefront amplitude.

\subsection{Transport dynamics between two leads}

In \cref{ld2_sec}, we examine a setup where two 1D attractively interacting leads at half-filling are connected via a Kondo impurity, as illustrated in \cref{fig_mid_cur}(a), a configuration closely related to the realization of hybrid superfluid-magnetic impurity devices. This setup reveals ground state properties similar to the case of the 1D bath studied in \cref{sec_sg_12}. However, it features a $\pi$ (or 0) phase shift between the SC order parameters in the left and right leads in the doublet (singlet) phase~\cite{Franceschi2010}, which we attribute to the combined effects of fermionic statistics and Cooper pair scattering through the Kondo impurity.

The primary focus of this section is the analysis of transport dynamics under a bias voltage $V$ between the two leads. We explore the time evolution where the system, initially in its ground state (either in the doublet or singlet phase, depending on the system parameters), is subjected to the bias voltage at time $t=0$. The non-equilibrium evolution of the current and system order parameters results in several interesting transient phenomena and a rich transport phase diagram with several regimes, as summarized in \cref{fig_mid_cur}(b). We provide a systematic analysis of the regimes shown in \cref{fig_mid_cur}(b), beginning with regimes I, II, and III in the doublet phase, which are connected by smooth crossovers, and then analyzing regime IV in the singlet phase.

{\it Regime I.---} In this regime, where the Kondo coupling $J$ is significantly smaller than the lattice hopping $t_h$ and the attractive Hubbard interaction $U$, the Kondo impurity simply acts as a weak link between the SC baths. As expected, this leads to the standard AC Josephson effect for small bias voltages. The current oscillates periodically, with the AC current amplitude that gets larger as $J$ increases. Additionally, there is a DC component of the current, that is small at small bias voltages, but becomes appreciable when the voltage exceeds twice the SC gap ($V>2\Delta$), which is sufficient to break Cooper pairs, and the resulting DC component indicates the transport of individual quasiparticles.

{\it Regime II.---} Increasing either the Kondo coupling $J$ or the applied voltage $V$ causes the transport behavior to deviate from that in regime I. In regime II, the system exhibits non-trivial time-dependence of the current, following sudden application of the bias voltage. At short times we find an oscillating AC Josephson current. However, these oscillations gradually damp out due to a dynamical transition from SC to CDW order in the bath, which dynamically restores the $U(1)$ symmetry. This transition is driven by the strong competition between the SC and CDW fluctuations in the attractive Hubbard model at half-filling. Notably, the Kondo correlations near the impurity can be transiently enhanced during intermediate times, leading to a significant transient peak in the current. Overall, the transport dynamics in regime II reveal a rich interplay between charge, magnetic, and superconducting orders, characterized by multiple dynamical crossovers.

{\it Regime III.---} Regime III appears when one further increases either $J$ or $V$ compared to regime II. In regime III, we observe substantial AC and DC components of the current simultaneously. This feature is attributed to increased DC conductance through the impurity, resulting from the partial Kondo cloud. Higher DC conductance facilitates stronger DC transport, causing the charge densities in the two leads near the impurity site to deviate from half-filling. In the attractive Hubbard model, SC order is favored over CDW order away from half-filling~\cite{scalettar1989phase,Moreo1991}, which suppresses density fluctuation and impedes the transition into the CDW state.

{\it Regime IV.---} In regime IV, formation of the Kondo singlet screens the impurity's magnetization and enhances the DC conductance. Upon applying a bias voltage $V$, the current quickly reaches a steady state, where the steady current $I_s(V)$ follows Ohm's law, $I_s(V) = G(V) \cdot V$, with $G(V)$ as the steady-state DC conductance. We systematically study $G(V)$ as a function of $V$ across various parameters. In the benchmark case of a non-interacting bath, we observe the expected quantized zero-bias conductance $G(0) \approx 2e^2/h$ and a quadratic nonlinearity $(2e^2/h - G(V)) \propto V^2$ for small $V$~\cite{oguri2001fermi}. We then examine the impact of the attractive interaction $U$. We find two interesting phenomena: (a) $U$ can anomalously enhance the small-bias conductance to exceed the quantized value ($G(V \ll 1) > 2e^2/h$), and (b) the AC Josephson current is significantly reduced in this regime. Both phenomena have been experimentally observed~\cite{husmann2015connecting,krinner2016mapping,hata2018enhanced,watfa2021collapse}, although their microscopic mechanisms remain either debatable~\cite{PhysRevLett.82.4086,hata2018enhanced,kanasz2016anomalous,liu2017anomalous,uchino2017anomalous} for (a) or unknown for (b). Our further analysis reveals that (a) results from additional chemical potential differences induced by the attractive Hubbard interaction. Additionally, (b) arises from a dynamic reduction of the phase difference in the SC order parameters between the two leads. Our theory thus provides an alternative, explicit microscopic mechanisms underlying these phenomena.

We also briefly discuss the transport phase diagram at other fillings. In particular, the attractive Hubbard model at non-half-filling favors the SC order~\cite{scalettar1989phase,Moreo1991}. Consequently, the transport phase diagram in non-half-filling cases is expected to include only regimes I, III, and IV.

\subsection{Experimental consideration on ultracold atom platforms}

In \cref{sec_imple}, we discuss the potential realization of the physics discussed in this paper using ultracold atom platforms. The setups we studied can generally be realized by integrating existing technologies. Notably, the Kondo interaction can be realized using alkaline earth atoms, such as fermionic $^{173}\rm Yb$, in optical lattices~\cite{zhang2020controlling,scazza2014observation,cappellini2014direct,Riegger2018}. Furthermore, this platform inherently supports the Hubbard interaction, which can be made attractive by adjusting the magnetic field~\cite{hofrichter2016direct,zhang2020controlling,olshanii1998atomic,bergeman2003atom}. By utilizing single-site spin-resolved imaging in quantum gas microscopes~\cite{hilker2017revealing}, magnetic and charge correlations within the system can be measured. Additionally, the transport setup can be implemented using tailored optical lattice potentials combined with additional laser beams~\cite{Bloch2008a,husmann2015connecting}.

\section{Theoretical framework}
\label{sec_theo}
\subsection{The general Hamiltonian}
We consider a general scenario where a magnetic impurity coupled to an attractive Fermi-Hubbard bath of $N$ sites [cf.~\cref{fig0}(a,b) and \cref{fig_mid_cur}(a) for specific setups considered in this paper]. The system Hamiltonian
\begin{equation} \label{ham_gen}
H = {H_{{\rm bath}}} + {H_{\rm {int} }}
\end{equation}
includes the Hamiltonian
\begin{equation} \label{ham_bath}
{H}_{\text {bath}}= \sum_{{\bm j}{\bm l},\alpha\in (\uparrow, \downarrow) }h_{{\bm j}{\bm l}} c_{{\bm j}, \alpha}^{\dagger} c_{{\bm l}, \alpha } - U\sum\limits_{{\bm j}} c_{{\bm j}, \uparrow }^{\dagger} c_{{\bm j}, \downarrow }^{\dagger} c_{{\bm j}, \downarrow } c_{{\bm j}, \uparrow }
\end{equation}
 of the fermionic bath and the Kondo interaction
 \begin{equation} \label{ham_kondo}
\begin{aligned}
{H}_{\text{int}}=
\sum_{\gamma \in (x,y,z)} {\hat s^\gamma _{\text {imp}}}\cdot  \hat \Phi^\gamma
\end{aligned}
\end{equation}
of a spin-1/2 impurity to the host fermions~\cite{Kondo1964,hewson1997kondo}.

In~\cref{ham_bath}, $h_{{\bm j}{\bm l}}$ is the single-particle Hamiltonian, and $U$ is the strength of the attractive Hubbard interaction. To study various coupling geometries, in~\cref{ham_kondo} we introduce a bath operator
\begin{equation} \label{}
\hat \Phi^\gamma= \frac{1}{2} \sum_{\bm j \bm l, \alpha \beta \in (\uparrow, \downarrow) } J_{\bm j \bm l}^{\gamma}  c_{\bm j, \alpha }^{\dagger} \hat \sigma^\gamma_{\alpha \beta} c_{\bm l, \beta }
\end{equation}
to describe a general Kondo interaction between the impurity spin ${\hat s^{\gamma} _{\rm imp}} ={\hat \sigma^{\gamma} _{\rm imp}}/2$ and bath fermions at sites $\bm j$ and $\bm l$ with strength $J_{{\bm j}{\bm l}}^{\gamma}$, where ${\hat \sigma^{\gamma} _{\rm imp}}$ is the Pauli matrix.

The attractive Hubbard model at half-filling exhibits a global SU(2) pseudo-spin symmetry~\cite{yang1990so,zhang1990pseudospin}. Rotations in pseudo-spin space connect the SC ground state with the CDW ground state~\cite{scalettar1989phase,Moreo1991},
resulting in degenerate ground states. We demonstrate in \cref{sc_cdw_trans} that, for the various Kondo coupling geometries considered in this paper [\cref{para_1d,para_2d,ham_2lead}], the system Hamiltonian [\cref{ham_gen}] retains the SU(2) pseudo-spin symmetry. Therefore, we focus on cases where the bath exhibits SC order in the ground state, as results for the bath in the CDW phase can be derived through a pseudo-spin rotation transformation.

We generalize the non-Gaussian variational approach~\cite{Shi2018a} to investigate both the in- and out-of-equilibrium physics of a magnetic impurity immersed in an SC environment. 
The validity of this approach has been demonstrated through studies of the ground state and long-time dynamics in quantum impurity problems and superconducting systems~\cite{Ashida2018a,ashida2018variational, PhysRevX.11.041028}. We introduce this approach in the remainder of this section.

\subsection{Disentangling transformation}

The key idea of the variational ansatz is to find a canonical transformation that decouples the impurity from the bath degrees of freedom. To achieve this, we observe that the system has a parity symmetry, i.e., $[H,\hat P]=0$. The parity operator $\hat P = \hat \sigma _{\rm imp}^z{\hat P_{\rm bath}}$ ($\hat P^2=1$) consists of the impurity spin operator $\hat \sigma _{\rm imp}^z$ along the $z$-direction and the bath parity operator
\begin{equation} \label{pb}
{\hat P_{{\rm{bath}}}} = \left\{ {\begin{array}{*{20}{l}}
 {e^{i\frac{\pi }{2}( {{\hat N_ \uparrow } - {\hat N_ \downarrow }} )}},\quad  \textrm{in the even sector of } {\hat N},  \\
 i{e^{i\frac{\pi }{2}( {{\hat N_ \uparrow } - {\hat N_ \downarrow }} )}},\quad  \textrm{in the odd sector of } {\hat N},
\end{array}} \right.
\end{equation}
where ${\hat N={{\hat N_ \uparrow }+{\hat N_ \downarrow }}}$ is the total particle number operator in the bath and ${\hat N_ \alpha } = \sum\limits_{\bm j} n_{\bm j, \alpha}$ is defined by the number operator $n_{\bm j, \alpha} =  c_{\bm j,\alpha}^{\dagger} c_{\bm j,\alpha}$ of the spin-$\alpha$ particle at site $\bm j$.

We construct a unitary transformation~\cite{Ashida2018a, ashida2018variational}
\begin{equation} \label{transU}
\hat U_{\rm NGS} = \frac{1}{{\sqrt 2 }}\left( {1 + i{\hat \sigma} _{\rm imp}^y{{\hat P}_{{\rm{bath }}}}} \right)
\end{equation} 
that maps the parity operator $\hat P$ to an impurity operator as ${\hat U_{\rm NGS}^\dag }\hat P\hat U_{\rm NGS} = \hat \sigma _{{\rm{imp}}}^x$. Since $\hat P$ is a conserved quantum number in the original frame, the impurity operator $\sigma _{{\rm{imp}}}^x$ commutes with the Hamiltonian $H'=\hat U^{\dagger}_{\rm NGS}H \hat U_{\rm NGS}$ in the transformed frame, i.e., $[ {H',\sigma _{{\rm{imp}}}^x} ] ={\hat U_{\rm NGS}^\dag }[ {H,\hat P} ]\hat U_{\rm NGS} = 0$. This can also be verified by explicitly calculating the Hamiltonian $H'$ as $H'= H_{\rm bath} + H'_{\rm int}$, where the Kondo interaction term becomes
\begin{equation} \label{ham_trans}
H'_{\rm int} = \hat s_{{\rm{imp}}}^x{{\hat \Phi^x }} + {{\hat P}_{{\rm{bath}}}}\left( {\dfrac{{ - i{{\hat \Phi^y }}}}{2} + \hat s_{{\rm{imp}}}^x{{\hat \Phi^z }}} \right).
\end{equation}
Therefore, the spin operator $\hat s^x_{\rm imp}=\pm1/2$ along the $x$-direction becomes a classical variable in the transformed frame. Without loss of generality, we focus on the sector $\hat s^x_{\rm imp}=1/2$, which implies that the impurity is in the state $|+ \rangle_{\rm imp} \propto |\uparrow\rangle_{\rm imp} + |\downarrow\rangle_{\rm imp}$ in the transformed frame. The two sectors $\hat s^x_{\rm imp}=\pm1/2$ are simply related via a spin rotation along the $y$-axis.

Since the impurity degrees of freedom are completely decoupled in $H'$, we can approximate its ground state and non-equilibrium dynamics using the fermionic Gaussian state $| \psi_{\rm GS} (\Gamma_f)\rangle$~\cite{kraus2010generalized,weedbrook2012,Shi2018a}, characterized by its covariance matrix ${\Gamma_f} = {\langle {CC^\dag } \rangle _{\rm GS}}$. Here, the Nambu spinor $C= {( {c,{c^\dag }} )^T}$ and $c = \left( \{c_{ \bm j ,\uparrow }\}, \{c_{ \bm j ,\downarrow }\}  \right)$. Combining the transformation \cref{transU} with the Gaussian state, we obtain the variational ansatz
\begin{equation} \label{ngs_state}
|\Psi_{\rm NGS} ({\Gamma_f}) \rangle = \hat U_{\rm NGS}  |+ \rangle_{\rm imp}   |\psi_{\rm GS} ({\Gamma_f}) \rangle
\end{equation}
in the original frame, where the variational parameters are the $O(N^2)$ matrix elements of $\Gamma_f$.

We note that, in contrast to the variational ansatz~\cite{Ashida2018a,ashida2018variational}, which is inapplicable to spontaneously U(1)-symmetry-broken systems, the present ansatz~\cref{ngs_state} with $\hat P_{\rm bath}$ [\cref{pb}] is well-defined for baths with SC orders. As a paradigmatic example, the $s$-wave Bardeen-Cooper-Schrieffer (BCS) state~\cite{bardeen57b,tinkham2004introduction} is naturally an eigenstate of $\hat P_{\rm bath}$.

\subsection{Non-Gaussian variational approach for imaginary and real-time evolution}
\label{var_method_main}

To determine the variational parameters in $\Gamma_f$ for the ground state properties and out-of-equilibrium dynamics, we employ the time-dependent variational principle~\cite{kraus2010generalized,Shi2018a}. In imaginary- and real-time evolutions, $\Gamma_f$ obeys
\begin{equation} \label{ite_eom}
{d _\tau }{\Gamma_f} = \{ {{\cal H},{\Gamma_f}} \} - 2{\Gamma_f}{{\cal H}}{\Gamma_f},
\end{equation}
\begin{equation} \label{rte_eom}
{d _t } {{\Gamma_f}} =  - i[ {{{\cal H}},{\Gamma_f}} ],
\end{equation}
where the mean-field Hamiltonian is ${\cal H} = 2\delta E_{\rm NGS} / \delta \Gamma_f $, with the variational energy $E_{\rm NGS} = \langle \Psi_{\rm NGS} | H |  \Psi_{\rm NGS} \rangle$. We apply the generalized Wick's theorem~\cite{Shi2018a} to obtain $E_{\rm NGS}$ and ${\cal H}$. In particular, the Hubbard interaction term $c_{\bm j ,\uparrow}^{\dagger} c_{\bm j ,\downarrow}^{\dagger} c_{\bm j ,\downarrow} c_{\bm j ,\uparrow}$ in \cref{ham_trans} leads to quadratic terms in ${\cal H}$, as
\begin{equation} \label{main_wick}
\begin{aligned}
c_{\bm j\uparrow}^{\dagger} c_{\bm j\downarrow}^{\dagger} c_{\bm j\downarrow} c_{\bm j\uparrow} &\longrightarrow\langle c_{\bm j\uparrow}^{\dagger} c_{\bm j\downarrow}^{\dagger}\rangle c_{\bm j\downarrow} c_{\bm j\uparrow}-\langle c_{\bm j\uparrow}^{\dagger} c_{\bm j\downarrow}\rangle c_{\bm j\downarrow}^{\dagger} c_{\bm j\uparrow} \\
&+\langle n_{\bm j, \uparrow} \rangle n_{\bm j, \downarrow}+ {\rm H.c.}
\end{aligned}
\end{equation}

The detailed procedure for solving \cref{ite_eom,rte_eom}, including the calculation of ${\cal H}$ and $E_{\rm NGS}$, is presented in \cref{mod_details}. The covariance matrix of the variational ground state is obtained via \cref{ite_eom} in the asymptotic limit $\tau \rightarrow \infty$, while the real-time dynamics can be studied using \cref{rte_eom}.

\section{Ground state properties}
\label{sec_sg_12}

When a magnetic impurity couples to an $s$-wave superconductor, the nature of the ground state is determined by the competition between the SC order, characterized by the SC gap $\Delta$, and the Kondo interaction, characterized by the Kondo temperature $T_K$~\cite{balatsky2006impurity}. When $T_K\ll \Delta$, electrons in the bath form Cooper pairs, and the impurity spin behaves as an almost free spin-$\frac{1}{2}$. Consequently, the ground state of the entire system is doubly degenerate, with a total spin of $\frac{1}{2}$, referred to as the \textit{doublet} phase. In the opposite limit, $T_K \gg \Delta$, the magnetic impurity is screened by Bogoliubov excitations on top of the BCS state, and a Yu-Shiba-Rusinov bound state~\cite{yu1965bound,shiba1968classical,rusinov1969superconductivity,rusinov1969theory} forms, referred to as the \textit{singlet} phase. Generally, as $T_K/\Delta$ varies, a first-order transition between the singlet and doublet phases~\cite{soda1967s,Satori1992,balatsky2006impurity,huang2024large} is expected [cf.~\cref{fig0}(c)].

In this section, we study the ground state properties of a Kondo impurity coupled to a 1D or 2D bath, focusing on the aforementioned singlet-doublet phase transition. For the remainder of this paper, we will primarily use the Kondo coupling $J$ and the attractive Hubbard interaction $U$ to characterize the system, instead of $T_K$ and $\Delta$. The relationship between $T_K$ and $\Delta$ as functions of $J$ and $U$ is shown in \cref{apd_1d} for reference.

\subsection{Setup}

In this section, we consider two different bath geometries. In the first geometry [cf.~\cref{fig0}(a)], the impurity is coupled to the central site $j=0$ of a 1D bath with a length of $N=2L+1$. In this case, the single-particle term and the Kondo coupling term in \cref{ham_bath,ham_kondo} are given by
\begin{align} \label{para_1d}
	h_{jl}^{\rm 1D} &= -t_h (\delta_{j,l-1} + \delta_{j,l+1}) -\mu \delta_{j,l} ,\nonumber \\
	J_{jl}^{\rm 1D} &= J \delta_{j,0}\delta_{l,0},
\end{align}
where $t_h$ is the nearest-neighbor hopping strength and $\mu$ is the chemical potential. Hereafter, we set $t_h = 1$ as the unit and focus on the antiferromagnetic Kondo coupling~\cite{Kondo1964,hewson1997kondo}, i.e., $J>0$.

In the second geometry [cf.~\cref{fig0}(b)], the impurity is coupled to the central site ${\bm j}=(0,0)$ of a 2D square lattice with a size of $N=(2L+1)\times (2L+1)$. The single-particle term and the Kondo coupling term are given by
\begin{align} \label{para_2d}
	h_{\bm j \bm l}^{\rm 2D} &= -t_h (\delta_{j_x,l_x-1}\delta_{j_y,l_y} + \delta_{j_x,l_x+1}\delta_{j_y,l_y}  \nonumber \\
	&+ \delta_{j_x,l_x}\delta_{j_y,l_y-1} + \delta_{j_x,l_x}\delta_{j_y,l_y+1}) - \mu \delta_{j_x,l_x}\delta_{j_y,l_y},  \nonumber \\	
	J_{\bm j \bm l}^{\rm 2D} &= J \delta_{\bm j,\bm 0}\delta_{\bm l,\bm 0},
\end{align}
where $\bm j \equiv (j_x,j_y)$ and $\bm l \equiv (l_x,l_y)$, with each index taking an integer value between $-L$ and $L$.

\begin{figure*}
	\centering
	\includegraphics[width=0.9\textwidth]{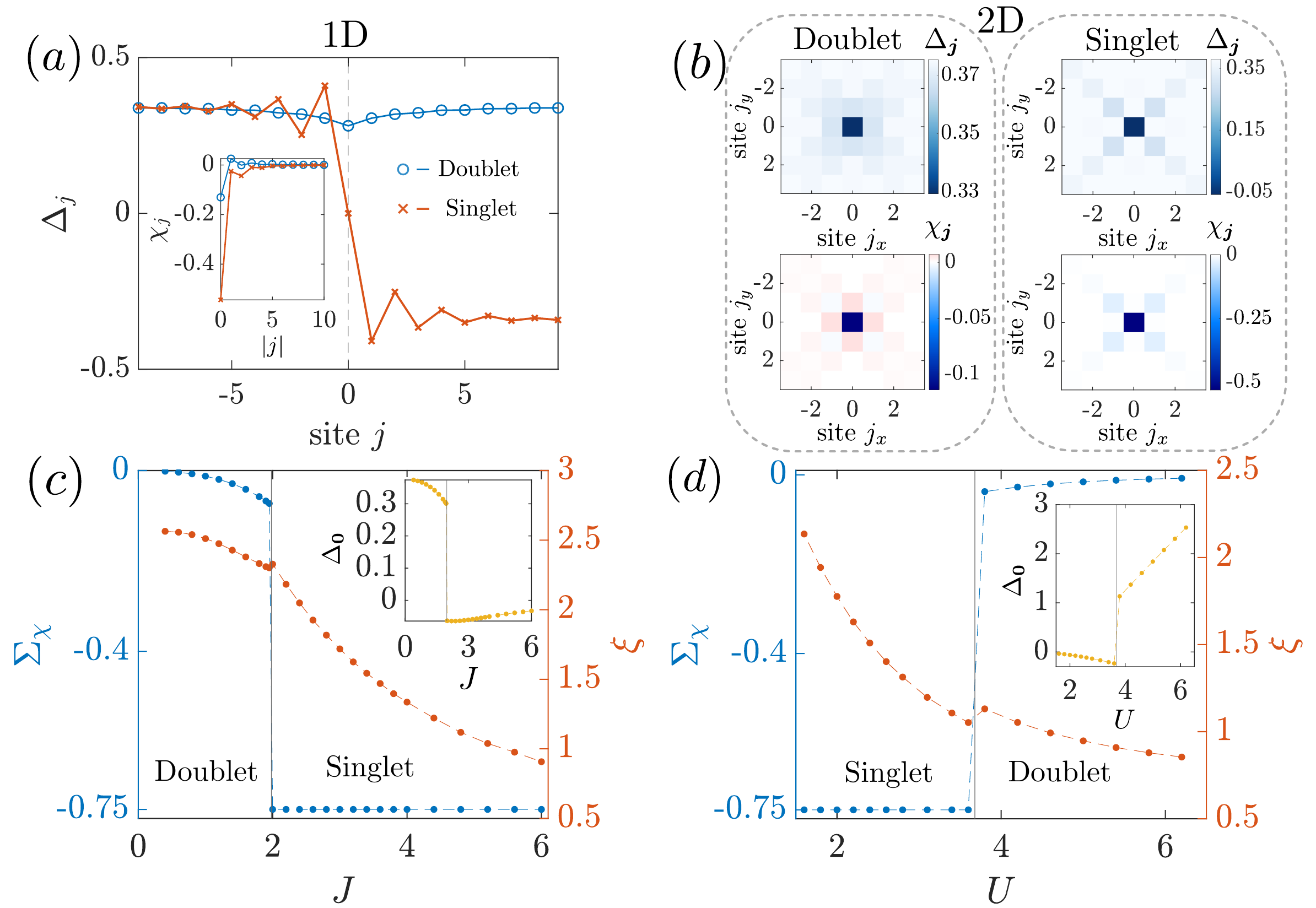}
        \caption{The behavior of order parameters in the ground state. (a) The SC order parameter $\Delta_j$ on the 1D bath (system size $N=401$) near the impurity for both the doublet phase $(U=2.0,J=1.6)$ and the singlet phase $(U=2.0,J=3.2)$, where the impurity site $(j=0)$ is marked by the dashed vertical line. The inset shows the spatial distribution of the impurity-bath spin correlation $\chi_j$. (b) The gap and impurity-bath spin correlation near the Kondo impurity, which is coupled to the center ${\bm j}  = (0,0)$ of the 2D bath with system size $N=41\times 41$, for both the doublet phase $(U=2.0,J=1.6)$ and the singlet phase $(U=2.0,J=3.2)$. (c) The sum of the impurity-bath spin correlation $\Sigma_{\chi}$ [cf.~\cref{sum_chi}] and the size of the (partial) Kondo cloud $\xi$ [cf.~\cref{ck_var}] as a function of the Kondo coupling $J$ for the 2D bath, with the Hubbard interaction $U=2.0$. The vertical line marks the singlet-doublet phase transition point. The inset shows the SC order parameter $\Delta_{\bm 0}$ at the lattice site ${\bm 0}=(0,0)$ that couples to the impurity. (d) The same as (c), but as a function of the Hubbard interaction strength $U$, with a fixed Kondo coupling $J=3.0$.}
        \label{fig1}
\end{figure*}

\subsection{Singlet-Doublet phase transition}
\label{sd_trans}

We investigate the singlet-doublet phase transition in the system described by~\cref{para_1d,para_2d} by numerically solving the corresponding imaginary time evolution~\cref{ite_eom} for the half-filling case, considering different total spins ${{\hat S}^2} =  \sum\limits_\gamma  {{{( {\hat \sigma _{{\rm{imp}}}^\gamma  + \sum\limits_{ {\bm j},\alpha \beta } {c_{{\bm j},\alpha }^{\dagger}\sigma _{\alpha \beta }^\gamma {c_{{\bm j},\beta }}} } )}^2}}$ in the doublet phase ($\hat S^2=3/4$) and the singlet phase ($\hat S^2=0$). By comparing the ground state energies $E_d$ in the doublet phase and $E_s$ in the singlet phase, we determine the true ground state of the system.

Figure~\ref{fig0}(d,e) presents the energy difference ${E_s} - {E_d}$ between the ground states in the singlet and doublet phases as a function of the Kondo coupling strength $J$ for the 1D and 2D systems. Here, ${E_s} - {E_d}$ is plotted in units of the SC gap $\Delta$, extracted from the bulk value of the SC order parameter 
\begin{equation} \label{}
\Delta_{\bm j}= U\langle c_{{\bm j}, \downarrow} c_{{\bm j}, \uparrow} \rangle	
\end{equation}
far from the impurity. As $J$ increases, the sign change of ${E_s} - {E_d}$ from positive to negative indicates an energy level crossing, marking a first-order transition from the doublet phase to the singlet phase. Furthermore, the transition point $J_c$ (where ${E_s}={E_d}$) shifts to larger values as the Hubbard interaction strength $U$ increases [see inset of \cref{fig0}(d,e)], indicating that the phase transition is driven by the competition between the Kondo interaction and the attractive Hubbard interaction. For all parameter ranges, $|{E_s} - {E_d}| < \Delta$, showing that the singlet (doublet) excited state lies within the SC gap above the doublet (singlet) ground state.

\subsection{Behavior of order parameters}
\label{sec_gs_order}
The magnetic tendency induced by the impurity competes with the bulk SC order, leading to intertwined orders near the impurity and spatial variation in the order parameters. This behavior is characterized by the spatially dependent SC order parameter $\Delta_{\bm j}$ and the impurity-bath spin correlation function
\begin{equation} \label{}
{\chi_{\bm j}} =\sum_{\gamma} \langle {{{\hat \sigma }_{\rm imp}}^{\gamma} \cdot {{\hat \sigma }_{\bm j}^{\gamma}}} \rangle /4.	
\end{equation}
The formation of the Kondo singlet is characterized by the correlation between the impurity spin and all electrons in the bath as
\begin{equation} \label{sum_chi}
{\Sigma _\chi } \equiv \sum\limits_{\bm j} {{\chi_{\bm j}}},
\end{equation}
which saturates to ${\Sigma _\chi }=-3/4$ in the singlet phase~\cite{hewson1997kondo}.

In \cref{fig1}(a), the spatial distributions of the SC order parameter $\Delta_{\bm j}$ and the Kondo correlation ${\chi_{\bm j}}$ near the impurity are shown for the 1D bath of size $N=401$. For $(U=2.0,J=1.6)$ in the doublet phase, $\Delta_{\bm j}$ slightly decreases near the impurity, and a small amplitude ${\chi_{\bm j}}$ emerges (${\Sigma _\chi } \approx -0.045$), indicating locally suppressed SC order and the establishment of minimal Kondo correlations, i.e., the formation of a partial Kondo cloud~\cite{moca2021kondo}. In contrast, the spatial distributions of the order parameters are markedly different in the singlet phase. Here, with the formation of the Kondo singlet, ${\chi_{\bm j}}$ is significantly enhanced close to the impurity, and ${\Sigma _\chi }=-3/4$. Notably, $\Delta_{\bm j}$ is suppressed to zero near the impurity and changes sign when crossing the impurity, indicating a $\pi$ phase shift of the Cooper pair scattering off the Kondo singlet.

This $\pi$ phase shift arises from the scattering of a Cooper pair through a Kondo singlet. According to Friedel's sum rule, a single electron acquires a $\pi/2$ phase shift when scattered by a Kondo singlet~\cite{langreth1966friedel,nozieres1974fermi}. Consequently, when a Cooper pair, composed of two electrons, transmits from one side of the Kondo singlet to the other, it acquires a total $\pi$ phase shift. As a result, the order parameter exhibits a $\pi$ phase shift on opposite sides of the impurity. We emphasize that this $\pi$ phase shift in the bath naturally arises when the SC order parameter is self-consistently determined, which does not occur when the back-action of the impurity on the bath is ignored in the BCS model with a fixed SC order parameter~\cite{balatsky2006impurity}.

We also illustrate the behavior of the gap $\Delta_{\bm j}$ and the Kondo correlation ${\chi_{\bm j}}$ for the 2D bath [cf.~\cref{fig1}(b)] with a system size $N=41\times41$. In the doublet phase, the SC order parameter is slightly suppressed near the impurity, with a small amount of Kondo correlation established, similar to the 1D case [cf.~\cref{fig1}(a)]. In the singlet phase, however, the magnitude of the SC order parameter is significantly reduced near the impurity, accompanied by a sign change of the gap at the impurity site. When the Kondo singlet forms, the impurity captures an electron at the impurity site, leading to a strong suppression of the SC order parameter there, even causing it to become negative~\cite{Flatte1997,salkola1997spectral}.

\begin{figure*}
	\centering
	\includegraphics[width=1\textwidth]{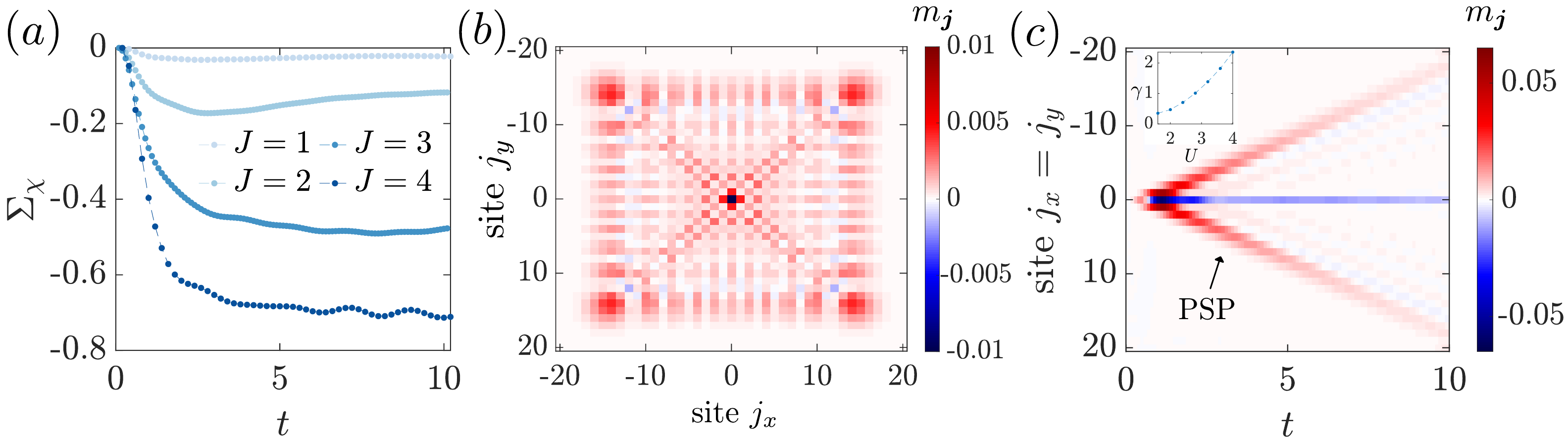}
        \caption{Relaxation dynamics of an impurity coupled to a 2D bath with a system size of $N=41 \times 41$. At time $t=0$, the Kondo coupling $J$ is turned on. (a) Evolution of the total impurity-bath spin correlation $\Sigma_\chi$ for various Kondo coupling values $J$ at a fixed attractive Hubbard interaction strength $U=2.0$. (b) Bath magnetization $m_{\bm j}$ of the 2D bath at time $t = 8.0$ during the quench dynamics for $U=1.6$ and $J=4.0$. (c) Evolution of the bath magnetization on diagonal sites of the lattice for $U=1.6$ and $J=4.0$. The inset shows the decay rate $\gamma$ of the PSP wavefront amplitude as a function of $U$.}
        \label{fig2}
\end{figure*}

To study the behavior of order parameters as functions of $J$ and $U$, we consider the SC order parameter $\Delta_{\bm 0}$ at the impurity site and the impurity-bath spin correlation ${\Sigma _\chi }$. Additionally, the size of the (partial) Kondo cloud is captured by the Kondo screening length
\begin{equation} \label{ck_var}
	\xi = \sqrt { \frac{\sum_{\bm j} |\chi_{\bm j}| \cdot r_{\bm j}^2}{\sum_{\bm l} |\chi_{\bm l}|} }.
\end{equation}
Here, $r_{\bm j}$ denotes the distance from the lattice site ${\bm j}$ to the impurity site $\bm 0$. For a 1D bath and a 2D bath, $r_{j} = |j|$ and $\sqrt{j_x^2 + j_y^2}$, respectively. We focus on the 2D bath in the rest of this section; similar results for the 1D bath are left in \cref{apd_1d}.

In \cref{fig1}(c) we show $\Delta_{\bm 0}$ and ${\Sigma _\chi }$ as functions of $J$. In the doublet phase, the magnitude of ${\Sigma _\chi }$ is a monotonically increasing function of $J$, indicating the formation of a stronger partial Kondo cloud, with the expected scaling ${\Sigma _\chi } \sim J^2$~\cite{moca2021kondo}. The screening length $\xi$ of the Kondo cloud decreases as $J$ increases, indicating that the Kondo cloud becomes more localized. The buildup of the impurity-bath spin correlation also leads to a decrease in $\Delta_{\bm 0}$ at the impurity site.

The order parameter $\Delta_{\bm 0}$ and the spin correlation ${\Sigma _\chi }$ exhibit an abrupt jump when crossing the phase transition point. In the singlet phase, ${\Sigma _\chi }=-3/4$ indicates the formation of a perfect Kondo singlet, with the screening length $\xi$ decreasing as $J$ increases. The SC order parameter $\Delta_{\bm 0}$ at the impurity site changes sign in the singlet phase, and its amplitude $|\Delta_{\bm 0}|$ further decreases with increasing $J$ due to the suppression of the SC order by the Kondo coupling.

We present the same set of order parameters as functions of the attractive interaction strength $U$ in \cref{fig1}(d). In this case, the SC order is enhanced with increasing $U$, which reduces the screening length $\xi$ in both the singlet and doublet phases.

In short, the ground state properties in the doublet phase agree well with previous studies using the BCS model~\cite{Satori1992,balatsky2006impurity,moca2021kondo}; however, in the singlet phase, a novel $\pi$ phase shift of the SC order parameter emerges in the 1D chain [cf.~\cref{fig1}(a)]. This $\pi$ phase shift is expected due to the scattering of Cooper pairs by the Kondo singlet, highlighting the importance of accurately capturing the back-action of the impurity on the bath in the self-consistent non-Gaussian approach.

\section{2D relaxation dynamics}
\label{sec_2d_relax}

In this section, we focus on the relaxation dynamics of a Kondo impurity coupled to a 2D attractive Hubbard bath of size $N=41\times 41$. The significant entanglement generated during this process~\cite{calabrese2007entanglement,ashida2018variational}, combined with the high-dimensional nature of the system, renders these dynamics intractable to simulate using existing tensor network methods~\cite{paeckel2019time}. In contrast, our variational approach based on the non-Gaussian ansatz~\cref{ngs_state,rte_eom} offers an efficient tool for studying the long-time relaxation dynamics in this model.

\begin{figure*}
	\centering
	\includegraphics[width=0.98\textwidth]{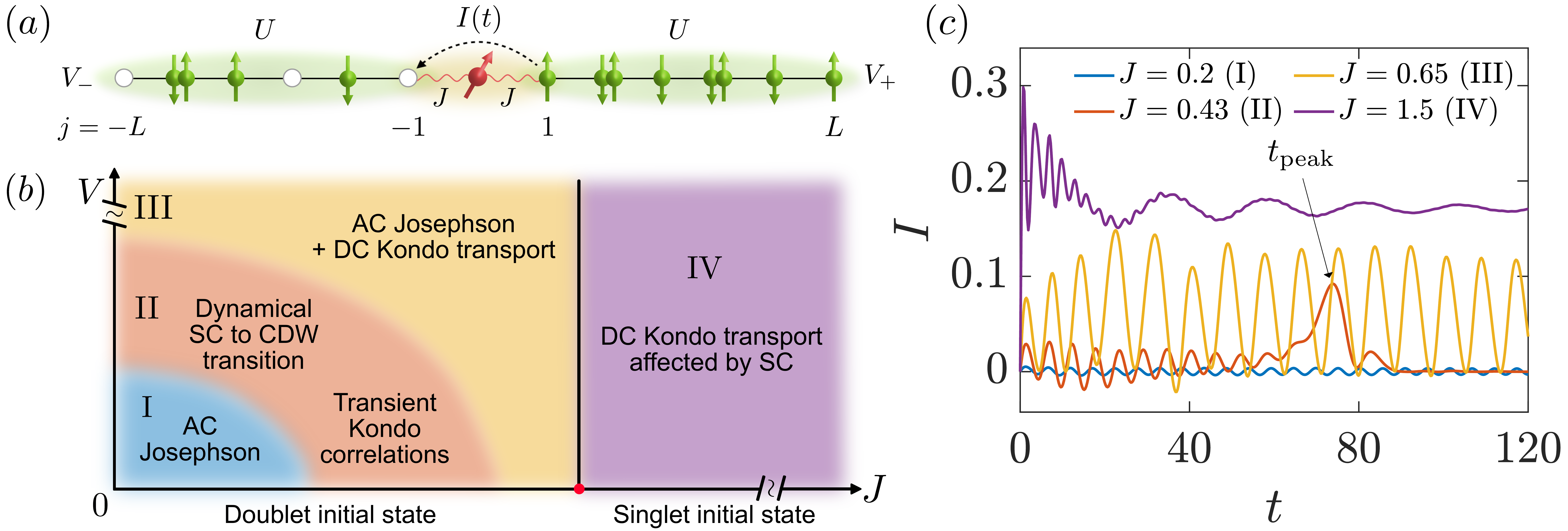}
        \caption{(a) In \cref{ld2_sec}, we examine two attractively interacting fermionic leads connected by a Kondo impurity [using the same notation as in \cref{fig0}(a)], where fermions can move between the leads via the Kondo coupling [cf.~\cref{ham_2lead}]. By applying external bias voltages $V_{\pm} = \pm V/2$, we study charge transport in this setup, focusing on the behavior of the current $I(t)$ (indicated by the curved dashed arrow). (b) Depending on the system parameters $(J,U,V)$, transient dynamics and transport characteristics reveal four qualitatively different regimes, that we mark as (I,II,III,IV). (c) The characteristic behavior of the charge current $I(t)$ in regimes I, II, III, and IV for various values of $J$. The other parameters are $U=2.0$, $V=0.6$, and $N=2L=400$. The transient current peak for $J=0.43$ (regime II) occurs at $t=t_{\rm peak}$ [see also~\cref{fig_mid_2}].}
        \label{fig_mid_cur}
\end{figure*}

We consider the system initially prepared in a product state between the impurity and the bath, $\left| {\Psi (0)} \right\rangle  = {\left|  \uparrow  \right\rangle _{{\rm{imp}}}}\left| {{\rm{SC}}} \right\rangle_{\rm 2D}$,
where the bath state $\left| {{\rm{SC}}} \right\rangle_{\rm 2D}$ represents the SC ground state at half-filling. At time $t=0$, the Kondo coupling $J$ is turned on, and the system evolves under the Hamiltonian \cref{ham_gen}.

Figure~\ref{fig2}(a) illustrates the evolution of the impurity-bath spin correlation $\Sigma_\chi(t)$ [\cref{sum_chi}]. The magnitudes of $\Sigma_\chi(t)$ quickly saturate to different plateau values for $J=1\sim4$ and $U=2$, indicating the rapid relaxation of the impurity toward forming a (partial) Kondo cloud. As $J$ increases, the magnitude of $\Sigma_\chi(t)$ relaxes to a larger value, with a near-perfect Kondo singlet forming ($\Sigma_\chi(t) \sim -3/4$) when $J \gtrsim U$.

The dynamical formation of the (partial) Kondo singlet also alters the bath magnetization, defined as 
\begin{equation} \label{}
m_{\bm j}=\frac{1}{2}\langle n_{{\bm j},\uparrow} - n_{{\bm j},\downarrow}\rangle.
\end{equation}
Figure~\ref{fig2}(b) shows a snapshot of $m_{\bm j}$ at $t=8$. A square-like wavefront is observed propagating in the bath, with a notable enhancement of the spin density wave along the diagonal lines. The square-like shape arises from the $C_4$ symmetry of the lattice, while the enhanced emission along the diagonal lines results from the divergent density of states at the bottom of the Bogoliubov particle band in the 2D bath~\cite{gonzaleztudela17a,gonzaleztudela17b,windt2024fermionic}.

In \cref{fig2}(c), we plot $m_{\bm j}$ at the diagonal sites over different times. A persistent sharp dip in $m_{\bm 0}$ at the impurity site $\bm{0}$ is observed, indicating the formation of a (partial) Kondo cloud. The remaining excitations form a pulse of spin polarization (PSP) propagating through the lattice, with the amplitude of the PSP wavefront decays exponentially as $\sim e^{-\gamma t}$ over time $t$ due to the finite SC gap [cf.~\cref{apd_2d_qu}]. As expected, the decay rate $\gamma$ increases as the SC gap becomes larger, corresponding to a stronger Hubbard interaction $U$, as shown in the inset of \cref{fig2}(c). It is also noted that the formation of the Kondo cloud and the propagation of the PSP, in turn, suppress the SC gap in the bulk [cf.~\cref{apd_2d_qu}].

In summary, the large-scale 2D dynamics demonstrate the competition between magnetism and superconductivity, influencing both the relaxation of the impurity and the propagation of excitations in the bath. Our results also show that directionally enhanced light or matter wave emission, previously studied in non-interacting systems where an impurity couples to structured reservoirs~\cite{gonzaleztudela17a,gonzaleztudela17b,windt2024fermionic}, also occurs in systems where both the impurity and the bath are strongly interacting.

\section{Transport dynamics between two SC leads}
\label{ld2_sec}

In this section, we present the intriguing transport phenomena in hybrid superfluid-magnetic impurity systems, as illustrated in \cref{fig_mid_cur}(a). Here, two 1D attractive SC leads (each of length $L$), described by the attractive Hubbard model, are coupled to the impurity through the Kondo interaction. Initially, the leads are prepared at half-filling, and the external voltages $V_{\pm}=\pm V/2$ are applied to drive the transport between the two leads.

The system Hamiltonian can be expressed in the same form as \cref{ham_bath,ham_kondo}, with the hopping and coupling matrix elements
\begin{align} \label{ham_2lead}
	h_{\eta j, \eta l}^{\rm 2leads} &= -t_h [\delta_{\eta j,\eta (l-1)} + \delta_{\eta j,\eta (l+1)}] +\left( eV_{\eta}- \mu\right)  \delta_{\eta j,\eta l}, \\
	J_{jl}^{\rm 2leads} &= J (\delta_{j,-1} \delta_{l,-1} + \delta_{j,-1} \delta_{l,1} + \delta_{j,1} \delta_{l,-1} + \delta_{j,1} \delta_{l,1}),		\nonumber
\end{align}
where $\eta \in (-,+)$ denotes the left and right leads, and the impurity couples to sites $j=1$ and $j=-1$.

\begin{figure}[h!]
	\centering
	\includegraphics[width=0.45\textwidth]{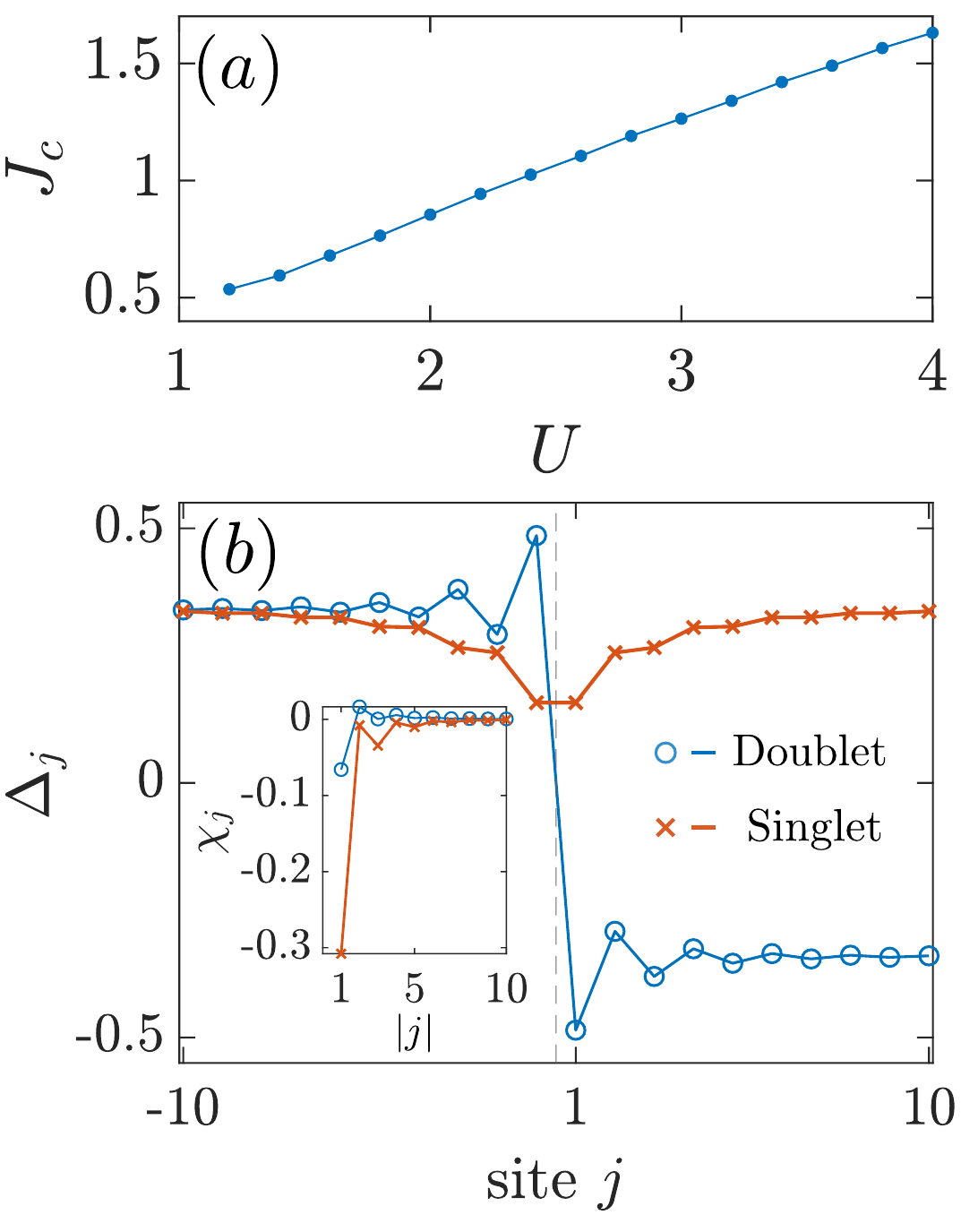}
        \caption{(a) The singlet-doublet transition point $J_c$ as a function of $U$ for the two-lead bath [cf.~\cref{fig_mid_cur}(a)] with a system size of $N=2L = 400$. (b) The SC order parameter $\Delta_j$ in the doublet phase $(U=2.0,J=0.6)$ and the singlet phase $(U=2.0,J=1.2)$ near the impurity location $j=0$ (indicated by the dashed vertical line). The inset shows the spatial distribution of the impurity-bath spin correlation $\chi_j$.}
        \label{ld2_sc_ph_fig}
\end{figure}

We first employ the non-Gaussian state ansatz~\cref{ngs_state} to obtain the ground state in the absence of a bias voltage. This setup exhibits ground state properties similar to the setup of a single lead [cf.~\cref{para_1d}], and we show the singlet-doublet transition point $J_c$ as a function of $U$ in \cref{ld2_sc_ph_fig}(a). A dual relation~\cite{shi2011two} between the phase distribution of the SC order parameter in two coupling geometries [cf.~\cref{para_1d,ham_2lead}] is displayed in \cref{fig1}(a) and \cref{ld2_sc_ph_fig}(b). More specifically, in the doublet (singlet) phase, there is a $\pi$ (0) phase shift between the SC order parameters in the left and right leads, in contrast to the 0 ($\pi$) phase shift observed in the single lead case [cf.~\cref{fig1}(a)]. This finding agrees with previous theoretical and experimental studies~\cite{Franceschi2010}. We attribute this effect to the interplay between fermionic statistics and the $\pi$ phase shift of the Cooper pair when scattered by the Kondo singlet [cf.~\cref{sec_gs_order}], which we elaborate on in \cref{apd_2l_gs}.

\begin{figure*}
	\centering
	\includegraphics[width=0.9\textwidth]{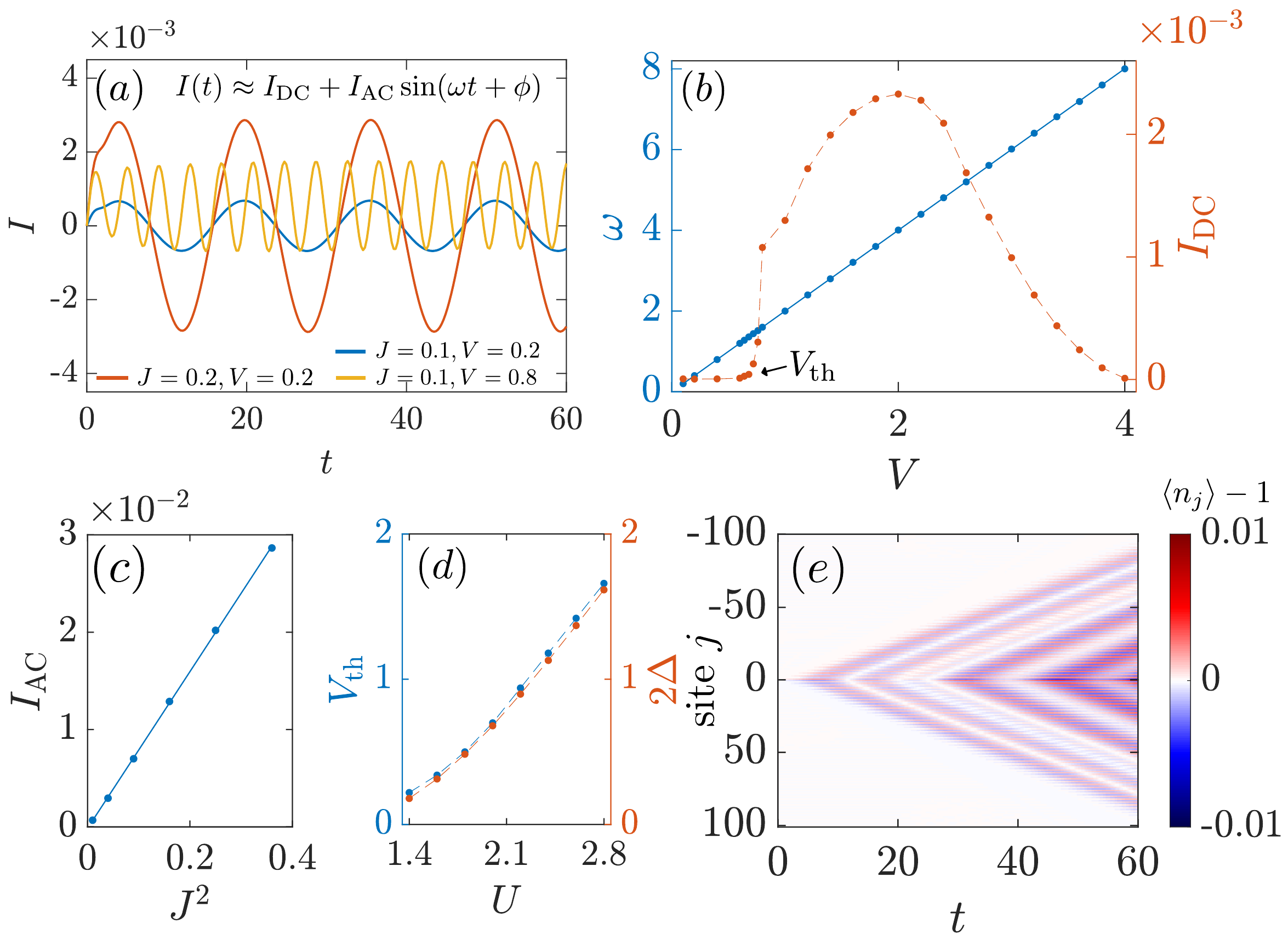}
        \caption{The AC Josephson effect in regime I [cf.~\cref{fig_mid_cur}(b)] of the two-lead transport dynamics (system size $N=2L=200$). (a) The current $I(t)$ for various Kondo coupling strengths $J$ and bias voltages $V$, with $U=2.0$. (b) The frequency of the AC component $\omega$ and the DC amplitude $I_{\rm DC}$ [cf.~\cref{ac_jj}] as functions of $V$, with $U=2.0$ and $J=0.1$. (c) The AC amplitude $I_{\rm AC}$ as a function of the effective tunneling strength $J^2$, with $U=2.0$ and $V=0.2$. (d) The threshold voltage $V_{\rm th}$ and twice the SC gap $2\Delta$ as functions of the Hubbard interaction $U$, with $J=0.1$. In (b-d), solid lines are linear fits, and dashed lines are visual guides. (e) The evolution of the charge density $\langle n_j \rangle$, with $U=2.0$, $J=0.1$, and $V=0.2$.}
        \label{fig3}
\end{figure*}

The remainder of this section is dedicated to analyzing transport dynamics under a bias voltage. We consider a time evolution in which the initial state $\left| {\Psi (0)} \right\rangle$ is the ground state of the system in the absence of a bias voltage. At time $t=0$, the bias voltages $V_{\pm} = \pm V/2$ are applied to drive the transport between the two leads, where the current
\begin{equation} \label{}
\begin{aligned}
		I(t)=i\frac{e J}{4 \hbar} \sum_{\alpha \beta}\left[\left\langle\hat{\boldsymbol{\sigma}}_{\mathrm{imp}} \cdot {c}_{+1, \alpha }^{\dagger} \boldsymbol{\sigma}_{\alpha \beta} {c}_{-1, \beta }\right\rangle-\text{ H.c. }\right].
\end{aligned}
\end{equation}
In the non-equilibrium state, the dynamics of the SC order parameter, charge transport, and spin correlation strongly depend on the system parameters, i.e., $(J,U,V)$, leading to a rich transport phase diagram illustrated in \cref{fig_mid_cur}(b). In this diagram, regimes I, II, and III in the doublet phase are connected by smooth crossovers, while regime IV lies in the singlet phase. Figure \ref{fig_mid_cur}(c) shows the representative behavior of the current $I(t)$ in regimes I, II, III, and IV. 

In the following, we systematically study the transport in the doublet phase in \cref{trans_db} and in the singlet phase in \cref{ld2_reg4}, and briefly discuss the change of the transport phase diagram at other fillings in \cref{phase_otherfill}.

\subsection{Transport in the doublet phase}
\label{trans_db}
In this subsection, we examine the transport dynamics in the doublet phase. We begin by analyzing regime I in \cref{ld2_reg1} and gradually increase the Kondo coupling $J$ to enter into regimes II and III [cf.~\cref{ld2_reg2,ld2_reg3}].

\begin{figure*}
	\centering
	\includegraphics[width=1\textwidth]{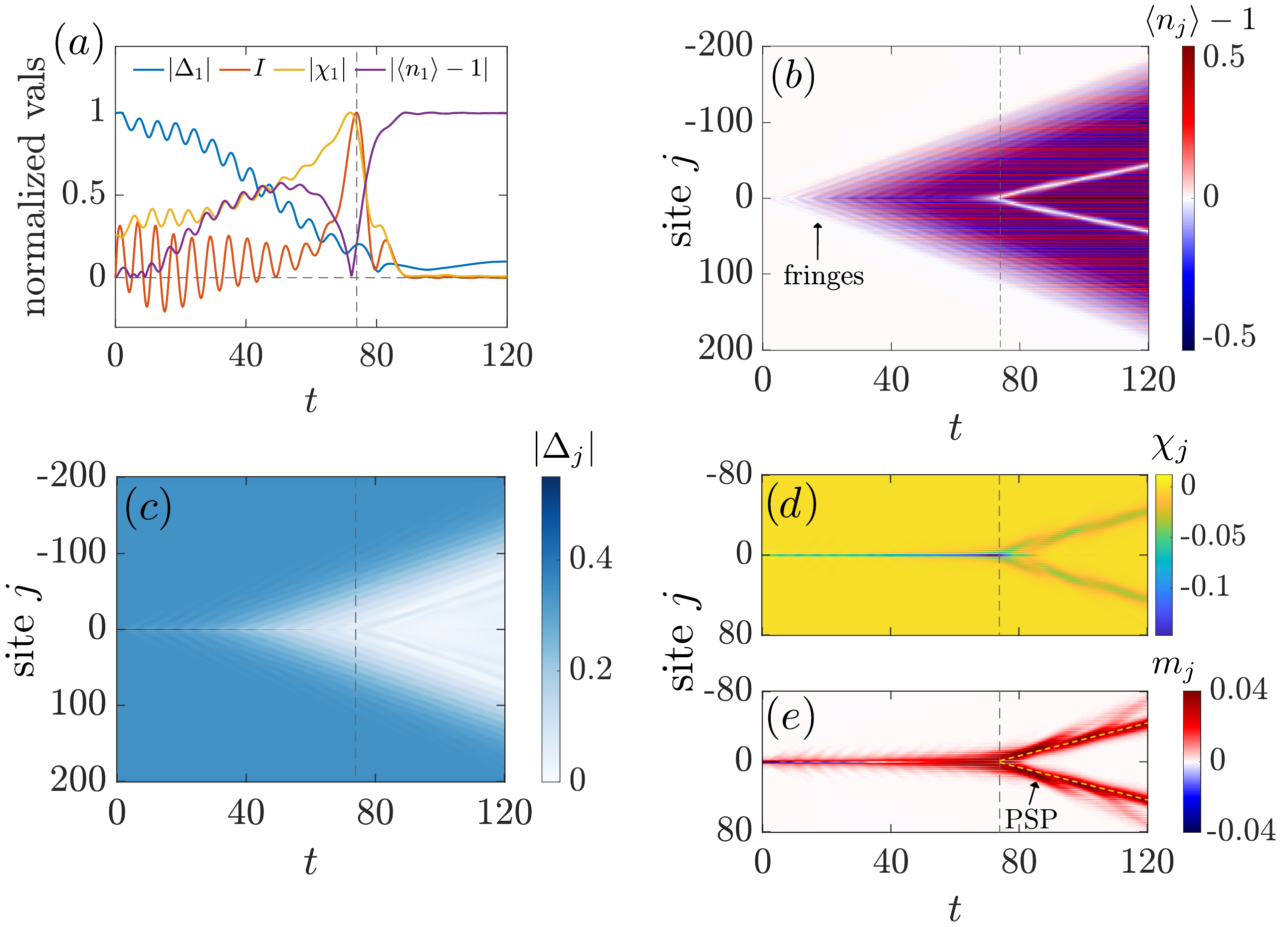}
        \caption{The dynamical transition from SC to CDW order in regime II [cf.~\cref{fig_mid_cur}(b)], with $U=2.0$, $V=0.6$, $J=0.43$, and $N=2L=400$ [(]the same parameters as the red curve in \cref{fig_mid_cur}(c)]. The time $t_{\rm peak}$, at which a current peak occurs in the red curve of \cref{fig_mid_cur}(c), is indicated by a dashed vertical line. (a) The evolution of the current $I$ and the order parameters at the contact point of the right lead ($j=1$), with values normalized to their maximum during the evolution. (b-e) Time evolution of (b) the charge density $\langle n_j \rangle$, (c) the amplitude of the SC order parameter $|\Delta_j|$, (d) the Kondo correlation $\chi_j$, and (e) the bath magnetization $m_j$. The tilted dashed line in (e) indicates the emitting PSP, which suppresses the CDW and SC order along its trajectory and leads to the tilted white stripe in (b).}
        \label{fig_mid_2}
\end{figure*}

\subsubsection{Regime I: AC Josephson effect}
\label{ld2_reg1}

We first consider the regime where the Kondo interaction $J$ is much smaller than both $t_h$ and $U$. In this case, the Kondo impurity functions as a weak link between the two SC baths. Consequently, under external DC bias voltages, the AC Josephson effect~\cite{josephson1962possible} is observed, as shown in \cref{fig3}(a). The current $I(t)$ can be expressed as
\begin{equation} \label{ac_jj}
I(t) \approx I_{\mathrm{DC}} + I_{\mathrm{AC}} \sin (\omega t + \phi_0)
\end{equation}
where $\omega$ is the frequency of the AC Josephson component, $I_{\rm AC}$ is its amplitude, $\phi_0$ is the initial phase, and $I_{\rm DC}$ is the DC amplitude. Figure~\ref{fig3}(a) shows that $I_{\rm AC}$ increases as the Kondo coupling $J$ strengthens, and the frequency $\omega$ increases with the bias voltage $V$. More specifically, in \cref{fig3}(b-c), we numerically verify the expected scalings $\omega = 2V$ and $I_{\rm AC} \propto J^2$ for the AC Josephson effect~\cite{josephson1962possible}. Figure \ref{fig3}(b) shows the DC amplitude $I_{\rm DC}$ as a function of $V$, which exhibits a sudden increase when $V$ exceeds a threshold voltage $V_{\rm th}$. In \cref{fig3}(d), we show that $V_{\rm th}$ is approximately equal to $2\Delta$ for various values of $U$. When $V>2\Delta$, a Cooper pair can break into two Bogoliubov quasiparticles, so this DC component is mainly carried by quasiparticles rather than Cooper pairs. In \cref{fig3}(e), we also illustrate the evolution of the charge density 
\begin{equation} \label{}
\langle n_{j} \rangle =  \langle n_{j, \uparrow} + n_{j, \downarrow} \rangle	
\end{equation}
in the two leads. Due to the oscillating current, charge density waves are periodically emitted from the two ends coupled to the impurity, and the direction of the net charge transport alternates periodically.

In regime I, Kondo physics does not play a crucial role, and our calculations accurately reproduce the expected AC Josephson effect.

\subsubsection{regime II: dynamical CDW to SC transition}
\label{ld2_reg2}

As $J$ increases from regime I, the system enters into regime II, with the typical behavior of the current $I(t)$ illustrated by the red curve in \cref{fig_mid_cur}(c). Initially, an AC Josephson current is established, which gradually dampens with a transient current peak emerging at an intermediate time $t_{\rm peak}$. Eventually, the current diminishes. This rich long-time behavior arises from the competition among SC order, charge fluctuations, and Kondo correlations at different timescales, as illustrated by typical behaviors of the normalized current and the order parameters around the impurity in \cref{fig_mid_2}(a). In the following, we systematically examine these behaviors.

\textit{(1) Initial evolution.}--- To understand the initial damping of the AC Josephson current, we examine the evolution of the charge density and the magnitude of the SC order parameter in \cref{fig_mid_2}(b,c). In the short term, the AC Josephson current induces oscillating density fringes without net charge transport, similar to what is observed in regime I [cf.~\cref{fig3}(e)]. Over longer evolution times, these oscillating density fringes induce CDW order around the impurity, with its size growing linearly over time, accompanied by the suppression of the SC order parameter in the same region. Because the CDW and SC ground states of the attractive Hubbard model at half-filling are degenerate, a small charge density fluctuation is sufficient to disrupt the SC order, inducing a transition to the CDW phase. The suppression of the SC order parameter thus leads to the observed reduction of the AC Josephson current.

\textit{(2) Dynamics around $t_{\rm peak}$.}--- Next, we analyze the transient current peak occurring at the intermediate time $t_{\rm peak}$. To this end, in \cref{fig_mid_2}(d,e), we plot the Kondo correlation $\chi_j$ and magnetization $m_j$ localized around the impurity, which characterize the partial Kondo cloud. During the time evolution, $|\chi_j|$ around the impurity gradually increases, reaching its maximal value at $t_{\rm peak}$, coinciding with the occurrence of the current peak. As shown in \cref{fig_mid_2}(a), the transient peak results from dynamical competitions among intertwined orders, where the Kondo correlations locally dominate the SC and CDW orders, significantly enhances the DC conductance~\cite{hewson1997kondo,pustilnik2004kondo} and leads to the current peak at $t_{\rm peak}$.

\textit{(3) Evolution toward final state.}--- Eventually, the transient partial Kondo cloud is destroyed, accompanied by the emission of a PSP [cf.~\cref{fig_mid_2}(c,d)], and the CDW phase forms throughout the baths [cf.~\cref{fig_mid_2}(a)]. Along the trajectory of the PSP, CDW order is locally suppressed [cf.~two white stripes in~\cref{fig_mid_2}(a)]. Here, the partial Kondo cloud forms only during the transient processes, as ground states of the attractive Hubbard bath only permit SC and CDW order. In the steady state, the system is dominated by CDW order with strongly suppressed SC order and Kondo correlations near the impurity, resulting in an almost zero current [cf.~the red curve in \cref{fig_mid_cur}(c)]. 

Overall, the dynamics in regime II exhibit rich interplay between intertwined orders. Notably, the dynamical SC to CDW transition restores the $U(1)$ symmetry, providing an example of the dynamical symmetry restoration~\cite{fagotti2014relaxation,piroli2016exact} in the quench dynamics of an interacting fermionic system.

\begin{figure*}
	\centering
	\includegraphics[width=1\textwidth]{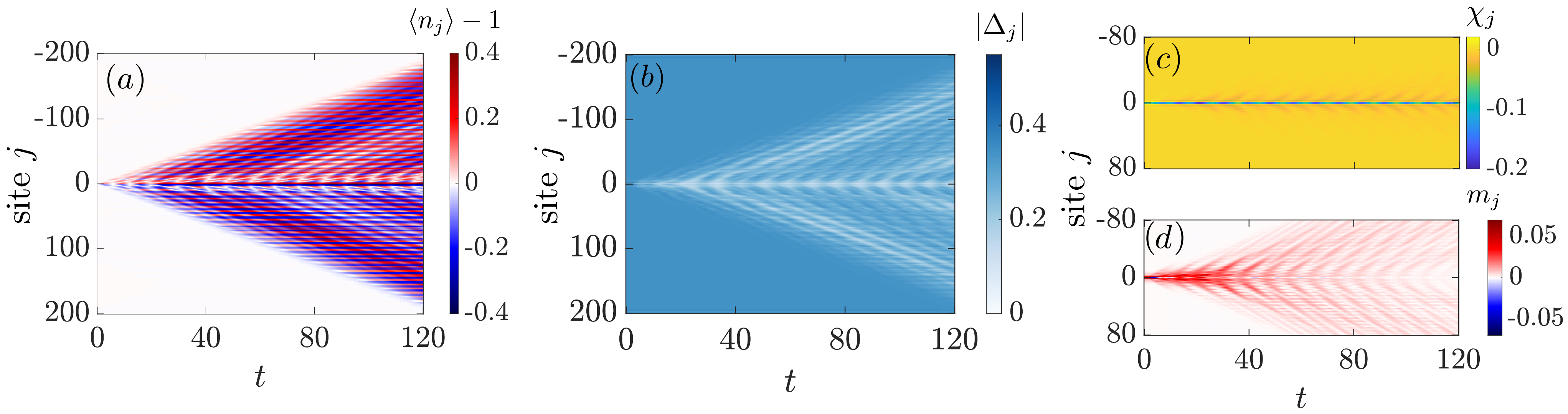}
        \caption{The behavior of order parameters in regime III [cf.~\cref{fig_mid_cur}(b)], with parameters $U=2.0$, $V=0.6$, $J=0.65$, and $N=2L=400$ [corresponding to the yellow curve in \cref{fig_mid_cur}(c)]. (a-d) The evolution of (a) the spatial charge density $\langle n_j \rangle$, (b) the amplitude of the SC order parameter $|\Delta_j|$, (c) the Kondo correlation $\chi_j$, and (d) the bath magnetization $m_j$.}
        \label{fig_mid_3}
\end{figure*}

\subsubsection{regime III: stable AC + DC transport}
\label{ld2_reg3}

By further increasing $J$, the system enters into regime III, where the typical current is depicted by the yellow curve in \cref{fig_mid_cur}(c). The current contains both significant AC and DC components, with the AC component remaining stable for a longer duration compared to regime II. The evolution of the density distribution in the bath shows a charge modulation (the stripe in \cref{fig_mid_3}(a)) induced by the AC component and a net charge transport (the inset in \cref{fig_mid_3}) driven by the DC component. As shown in~\cref{fig_mid_3}(b), the SC order parameter in this regime is maintained in contrast to regime II [cf.~\cref{fig_mid_2}(a,b)].

The behavior of the current can be understood as follows: As $J$ is considerable in regime III, the partial Kondo cloud forms with substantial spin correlations, facilitating the DC transport in the presence of the bias voltage~\cite{hewson1997kondo,pustilnik2004kondo}. Consequently, the two leads deviate from half-filling during the evolution [cf.~\cref{fig_mid_3}(a)]. Since the degeneracy of CDW and SC states is lifted away from half-filling, and the system favors the SC order~\cite{scalettar1989phase,Moreo1991}, the AC Josephson current is stabilized.

The evolution of the Kondo correlation in regime III is shown in \cref{fig_mid_3}(c,d). Compared to that in regime II [cf.~\cref{fig_mid_2}(c,d)], a partial Kondo cloud persists throughout the evolution, exhibiting small periodic oscillations induced by the AC Josephson current.

Overall, in the doublet phase, as $J$ varies, the system undergoes crossovers among three regimes [cf.~\cref{fig_mid_cur}(b)], characterized by the AC Josephson current in regime I, the dynamical SC to CDW transition with a transient Kondo peak in regime II, and the coexistence of DC and AC components due to dynamical stabilization of SC order in regime III. Finally, increasing the voltage $V$ also generally tunes the system from regime I to regime III [cf.~\cref{apd_cV_23}], thus leading to the transport phase diagram in the doublet phase [cf.~\cref{fig_mid_cur}(b)].

\subsection{Transport in the singlet phase}
\label{ld2_reg4}

For large $J > J_c$, the system enters into the singlet phase (regime IV), and the initial state $\left| {\Psi (0)} \right\rangle$ for time evolution is the ground state in the odd sector of $\hat N$~\footnote{It is also possible to choose $\left| {\Psi (0)} \right\rangle$ in the even sector of $\hat N$ when $J > J_c$. We expect the steady-state transport properties to be qualitatively the same, since for large $J > J_c$, the impurity can quickly relax and locally form a Kondo singlet, as observed in the 2D quench dynamics [cf.~\cref{fig2}(a)].}. In this scenario, the Kondo singlet forms, enhancing the DC conductance and facilitating charge transport even in the presence of the SC gap.

To systematically study the transport behaviors in this regime, we first examine the case of non-interacting leads in \cref{trans_reg4_noU} as a benchmark, then introduce attractive Hubbard interactions in \cref{trans_reg4_addU}. In \cref{trans_reg4_DC,trans_reg4_AC}, we further analyze two major phenomena in this regime: (1) anomalously enhanced small-bias DC conductance and (2) the suppression of the AC amplitude compared to that in the doublet phase. Our analysis reveals explicit microscopic mechanisms for these phenomena, providing insights into recent experimental observations~\cite{husmann2015connecting,krinner2016mapping,hata2018enhanced,watfa2021collapse}.

\begin{figure*}
	\centering
	\includegraphics[width=1\textwidth]{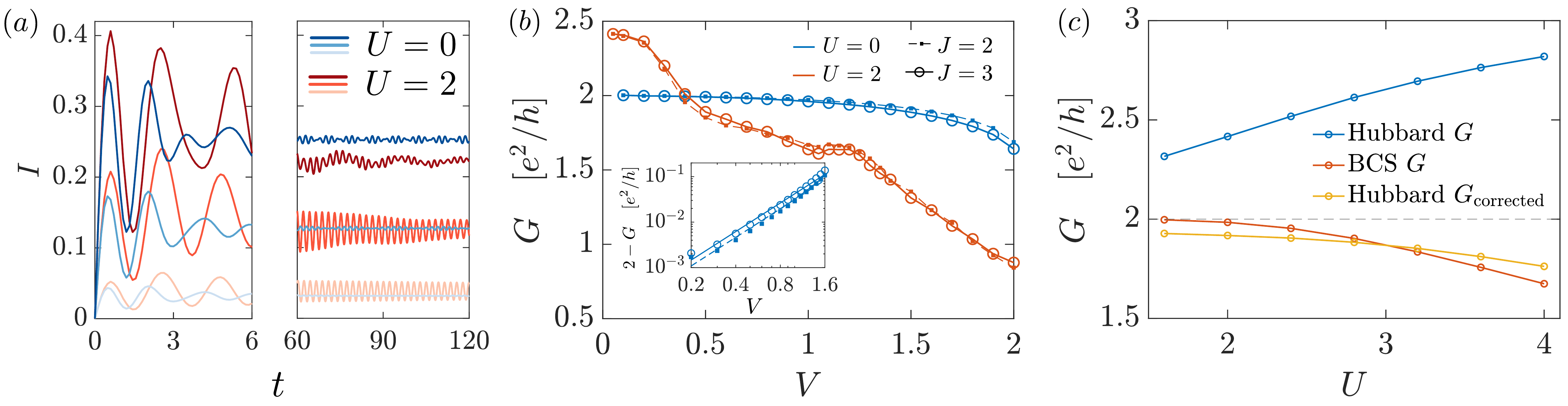}
        \caption{The two-lead transport dynamics in the singlet phase [regime IV in \cref{fig_mid_cur}(b)], studied with a system size of $N=2L=400$. (a) The current $I(t)$ for various bias voltages $V=0.1, 0.4, 0.8$ (colors from lighter to darker) for both non-interacting ($U=0$, red curves) and attractively interacting ($U=2$, blue curves) baths, with Kondo coupling $J=2$. (b) The steady-state DC conductance $G(V)$ as a function of $V$. The inset shows the deviation of $G(V)$ from the quantized value $2e^2/h$.(c) The small-bias ($V=0.01$) conductance $G(V=0.01)$ in the singlet phase ($J=2.5$) as a function of the Hubbard interaction $U$ for (i) Hubbard bath, (ii) BCS bath, and (iii) Hubbard bath with the corrected definition of the DC conductance $G_{\rm corrected}$ [\cref{G_cor}].}
        \label{regiv_fig}
\end{figure*}

\subsubsection{The case of non-interacting leads}
\label{trans_reg4_noU}
In the case of non-interacting leads ($U=0$), the system models a Kondo impurity coupled to two normal leads, and its transport properties have been extensively studied both theoretically and experimentally~\cite{pletyukhov2012nonequilibrium,antipov2016voltage,kretinin2011spin,kretinin2012universal,schwarz2018nonequilibrium,ashida2018variational,al2006adaptive,eckel2010comparative,bertrand2019reconstructing,schmidt2008transient,werner2010weak,thoenniss2023efficient,ng2023real}. In \cref{regiv_fig}(a), we present the time evolution of the current $I(t)$ for various bias voltages $V$, with $J=2$. In the linear-response regime ($t \lesssim 2$), the current develops an initial peak, which increases monotonically with $V$, in agreement with the Monte Carlo simulations for the Anderson model~\cite{antipov2016voltage}. The current then quickly reaches a steady-state value $I_s$ for $V=0.1, 0.4, 0.8$, with $I_s$ increasing with $V$, as expected from Ohm's law. In \cref{regiv_fig}(b), we plot the steady-state DC conductance $G(V) \equiv I_s / V$ as a function of the bias voltage $V$. We observe the quantized zero-bias conductance $G(0)\approx 2e^2/h$~\cite{pustilnik2004kondo,hewson1997kondo}, and confirm the quadratic behavior $(2e^2/h-G(V)) \sim V^2$ for small $V$~\cite{oguri2001fermi}.

\subsubsection{Adding attractive interaction}
\label{trans_reg4_addU}
When electrons are attractively interacting in the leads, the typical behaviors of the current in regime IV are shown in \cref{regiv_fig}(a) and the purple curve in \cref{fig_mid_cur}(c). The SC order in the system leads to both DC and AC components in the current $I(t)$, with the DC conductance $G(V)$ also shown in \cref{regiv_fig}(b). Compared to the case of non-interacting leads, the attractive interaction causes several major effects on $G(V)$:
\begin{enumerate}
    \item[{(1)}] For $V$ below a threshold value $V_c$, the SC order enhance $G(V)$, which can even exceed the quantized value, i.e., $G(V\ll 1) > 2e^2/h$.

    \item[{(2)}] For $V>V_c$, the presence of an attractive Hubbard interaction $U$ reduces $G(V)$ due to the suppression of Kondo correlations by the SC order. The competition between this reduction and the enhancement of $G(V)$ mentioned in (1) determines $V_c$.

    \item[{(3)}] In contrast to the monotonically decreasing $G(V)$ in the case of non-interacting leads, here $G(V)$ may slightly increase with $V$ in certain regimes [e.g., around $V \approx 1.2$ in \cref{regiv_fig}(b)]. This behavior arises from DC transport through single-particle channels when $V > 2\Delta$, as observed in \cref{fig3}(b).
\end{enumerate}

Among the above effects, the anomalously large DC conductance $G(V\ll 1) > 2e^2/h$ between two superfluids/superconductors is particularly intriguing, as this phenomenon has been observed in multiple experiments~\cite{husmann2015connecting,krinner2016mapping,hata2018enhanced}, with its microscopic mechanism under debate~\cite{PhysRevLett.82.4086,hata2018enhanced,kanasz2016anomalous,liu2017anomalous,uchino2017anomalous}. In \cref{trans_reg4_DC}, we present an explicit microscopic mechanism for this phenomenon in our setup.

Intuitively, the AC Josephson component typically increases with the tunneling rate between two leads via the Kondo coupling $J$, as shown in~\cref{fig_mid_cur}(c) for the doublet phase. However, suppression of the AC Josephson effect in the singlet phase (regime IV) is observed as $J$ increases from regime III [cf.~\cref{fig_mid_cur}(c)]. Notably, the suppression of the AC component was also experimentally observed in a carbon-nanotube (CNT)-based Josephson junction~\cite{watfa2021collapse}. Reference~\cite{watfa2021collapse} provides a phenomenological explanation for this observation; however, the underlying microscopic mechanism remains elusive. In \cref{trans_reg4_AC}, we also present an explicit microscopic mechanism for this phenomenon.

\subsubsection{Anomalous enhancement of small-bias DC conductance}
\label{trans_reg4_DC}

Here, we examine in detail the anomalous DC conductance $G(V \ll 1) > 2e^2/h$ observed in the case of an interacting bath [cf.~\cref{regiv_fig}(b)]. We compute the DC conductance $G(V=0.01)$ for a small bias $V=0.01$. For a fixed $J=2.5$, \cref{regiv_fig}(b) shows an almost linear increase of $G(V=0.01)$ with $U$. We explore the origin of the anomalous DC conductance by independently considering the contributions of the charge, spin, and pair fluctuations described by the Hartree, Fock, and Bogoliubov terms in \cref{main_wick}.

By retaining only the Bogoliubov terms $\langle c_{j,\uparrow}^{\dagger}c_{j,\downarrow}^{\dagger}\rangle c_{j,\downarrow}c_{j,\uparrow}+\rm{H.c.}$, the SC order parameter is self-consistently determined, while the charge and spin fluctuations are neglected.
We refer to this as the BCS bath. The behavior of $G(V=0.01)$ for the BCS bath is shown in \cref{regiv_fig}(c). We observe that $G(V=0.01) \approx 2e^2/h$ for small values of $U$, and it decreases monotonically with increasing attractive interaction strength. This indicates that the observed anomalous DC conductance $G(V \ll 1) > 2e^2/h$ is induced by charge or spin fluctuations. Moreover, since the system is spin-rotationally invariant in the initial state, the conservation of the total spin $\hat S^2=0$ implies that the system is spin-neutral during the transport ($\langle n_{j,\uparrow}\rangle = \langle n_{j,\downarrow}\rangle, \quad \forall j$), and the Fock terms $\langle c_{j\uparrow}^{\dagger} c_{j\downarrow}\rangle c_{j\downarrow}^{\dagger} c_{j\uparrow} + {\rm H.c}$ vanish. Inspired by this, we will study the effect of the charge fluctuations, described by Hartree terms $\langle n_{j, \uparrow} \rangle n_{j, \downarrow}+{\rm H.c.}
$ in \cref{main_wick}, on the anomalous enhancement of the DC conductance.

The DC amplitude between the two leads is primarily determined by the bias voltage $V$. In addition, the attractive interaction induces a correction 
\begin{equation} \label{eff_chem}
\delta \mu_{j}=U\langle n_{j,\alpha}\rangle,
\end{equation}
to the chemical potential through the Hartree terms [cf.~\cref{main_wick}]. Here, $\delta \mu_{j}$ depends on $U$ and the particle density at each site. This interaction-induced chemical potential is crucial during transport, as particle densities differ between the two leads.

Without loss of generality, we consider a bias voltage $V>0$, where fermions flow from the right to the left lead during transport. Assuming the system is initially half-filled, the time-dependent particle numbers at the contact points of the two leads (sites $j=-1$ and $j=1$) are expressed as:
\begin{equation} \label{}
\langle n_{\pm 1,\alpha} (t)\rangle  = \frac{1}{2} (1 \mp \delta n(t)),\qquad \alpha\in (\uparrow, \downarrow),
\end{equation}
where $\mp \delta n(t)$ ($\delta n(t)>0$) represents the time-dependent charge density deviation from half-filling at site $j=\pm 1$. The density difference $\delta n(t)$ results in a difference $\delta \mu_{-1}-\delta \mu_{1}=U\delta n(t)$ in the interaction-induced chemical potentials [\cref{eff_chem}] between the two leads, which effectively enhances the bias voltage $V$ to
 \begin{equation} \label{eff_bias}
eV'(t) = eV + U\delta n(t) > eV.
\end{equation}
As a result, when the system reaches the steady state, i.e., $\delta n(t) \rightarrow \delta n_{s}$ and $V'(t) \rightarrow V'_{s}$, the DC conductance increases by a factor of $eV'_{s}/eV = 1+U\delta n_{s}/eV$. This analysis agrees with the anomalous small-bias conductance $G(V\ll 1)>2e^2/h$ observed in \cref{regiv_fig}(b) and explains the linear increase of $G(V=0.01)$ with $U$ observed in \cref{regiv_fig}(c). To further substantiate this mechanism, we calculated a `corrected' DC conductance
\begin{equation} \label{G_cor}
G_{\rm corrected} \equiv I_s /V'_{s},
\end{equation}
which accounts for the interaction-induced chemical potential. Figure~\ref{regIV_AC_fig}(c) shows that $G_{\rm corrected}$ closely aligns with the results for the BCS bath. This agreement confirms that the anomalous DC conductance $G(V\ll 1) > 2e^2/h$ is primarily induced by charge fluctuations.

\subsubsection{Suppression of the AC Josephson current in the Kondo singlet regime}
\label{trans_reg4_AC}

\begin{figure}
	\centering
	\includegraphics[width=0.5\textwidth]{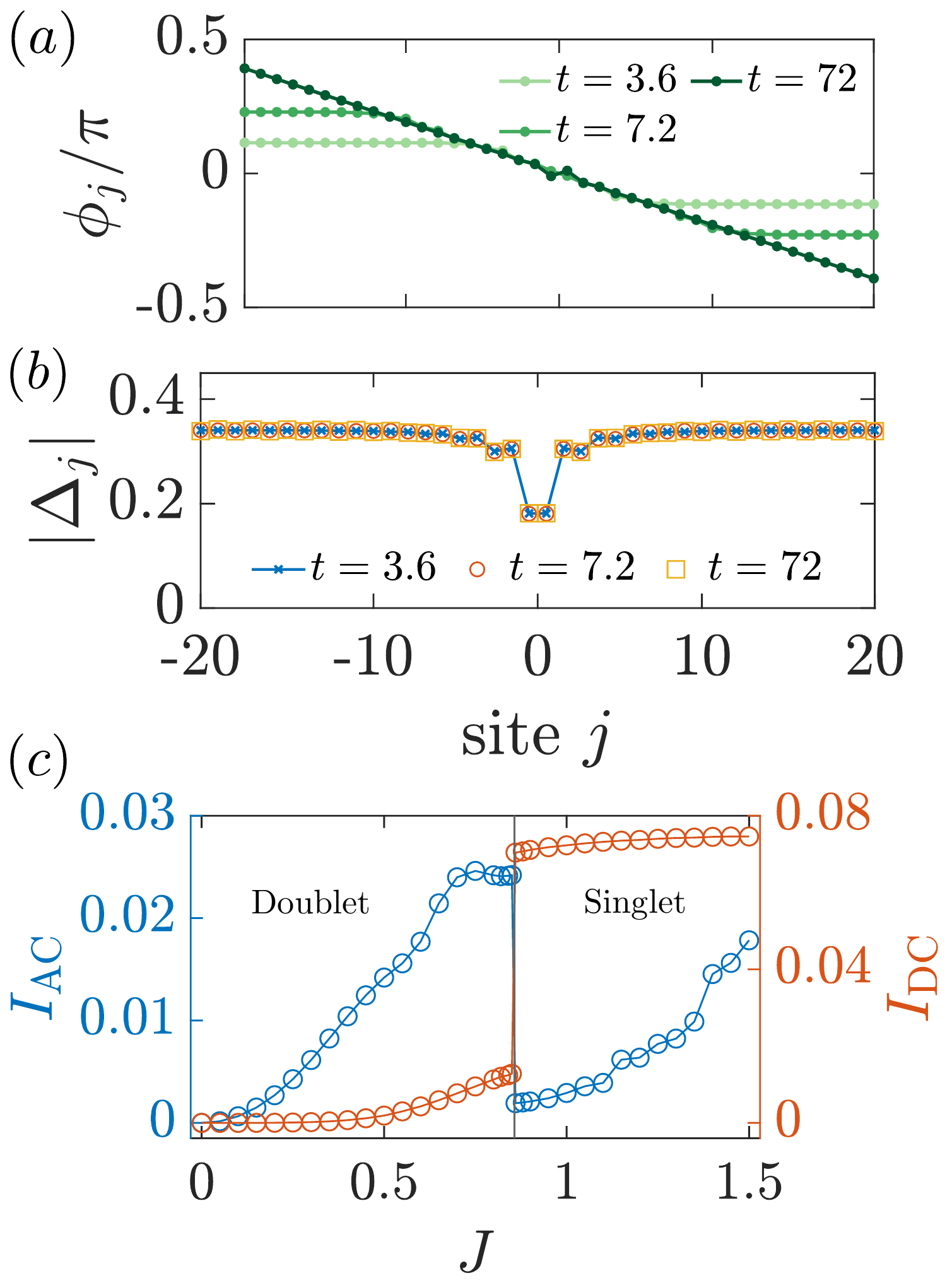}
        \caption{(a-b) The spatial distribution of (a) the phase $\phi_j(t)$ and (b) the amplitude $|\Delta_j(t)|$ of the SC order parameter at evolution times $t=3.6, 7.2, 72$, for $U=2.0$, $J=2.0$, and $V=0.1$. (c) The AC amplitude $I_{\rm AC}$ and the DC amplitude $I_{\rm DC}$ as functions of the Kondo interaction strength $J$, with $U=2.0$ and $V=0.2$. The vertical black line indicates the location of the singlet-doublet phase transition.}
        \label{regIV_AC_fig}
\end{figure}

We now analyze the suppression of the AC Josephson current in regime IV, as shown in \cref{fig_mid_cur}(c). According to the theory of the AC Josephson effect, the AC Josephson current scales as $I_{\rm AC}(t) \sim J^2 \sin^2(\delta \phi(t))$, where $\delta \phi(t) = \phi_{1}(t) - \phi_{-1}(t)$ is the phase difference of the SC order parameter between the contact points of the two leads. This implies that, for large $J$ in the singlet phase, the significant suppression of the AC Josephson current is related to the phase difference of the SC order parameter near the impurity. 

Figure~\ref{regIV_AC_fig}(a,b) shows the phase $\phi_j(t)$ and the amplitude $|\Delta_j(t)|$ of the SC order parameter at instants $t=3.6$, $7.2$ (initial times), and $72$ (final time). With the bias voltage $V$, one expects the phases of the SC order parameters in the bulk to change linearly with time, such that $\phi_+(t) = \phi_+(0) + Vt/2$ and $\phi_-(t) = \phi_-(0) - Vt/2$. This corresponds to the plateau regions away from the impurity in \cref{regiv_fig}(a) for $t=3.6$ and $7.2$. Remarkably, even for $V$ smaller than the SC gap, the strong DC conductance induced by the Kondo singlet leads to the DC transport of Cooper pairs through the multiple Andreev reflections~\cite{PhysRevLett.82.4086,nazarov2009quantum}, driving the super-current from the right to the left lead. The SC phase interpolated almost linearly between the two bulk values $\phi_\pm (t)$ indicates the formation of the super-current [cf.~\cref{regiv_fig}(a)] and a significantly reduced phase difference $\delta \phi(t)$ between the contact points. Since the amplitude $|\Delta_j(t)|$ of the SC order parameter near the impurity remains finite, as shown in \cref{regiv_fig}(b), the strong suppression of the AC Josephson current in the singlet phase is primarily due to the reduction in phase difference $\delta \phi(t)$.

To elucidate this phenomenon further, we present the DC and AC amplitudes $I_{\rm DC}$ and $I_{\rm AC}$ in \cref{regIV_AC_fig}(c) as functions of the Kondo interaction strength $J$, with $U=2$ and $V=0.2$. These amplitudes are obtained from 
the Fourier transformation of $I(t)$ in the steady state regime. Within each phase, $I_{\rm AC}$ increases with $J$ due to the enhanced Josephson tunneling rate. More importantly, as $J$ increases from the doublet to the singlet phase, there is a strong suppression of $I_{\rm AC}$ and an enhancement of $I_{\rm DC}$. Our theory naturally provides a microscopic mechanism for the key experimental finding in Ref.~\cite{watfa2021collapse}.

Overall, in the singlet phase, we observe strong charge transport characterized by a large DC amplitude and a suppressed AC component. We propose explicit microscopic mechanisms for the anomalous conductance $G(V\ll 1)>2e^2/h$ [cf.~\cref{trans_reg4_DC}] and the suppression of the AC component [cf.~\cref{trans_reg4_AC}]. The agreement between our theoretical predictions and experimental observations of these phenomena~\cite{hata2018enhanced,watfa2021collapse} supports the application of our theory for characterizing Kondo-SC systems in the singlet phase and demonstrates the necessity of a fully self-consistent approach that incorporates both charge and pair fluctuations.

\subsection{The potential change of the phase diagram in other fillings}
\label{phase_otherfill}
Changes in the filling factor impact the density of states at the Fermi surface. More importantly, away from half-filling, the degeneracy between CDW and SC is lifted, and the system favors the SC phase in the ground state~\cite{scalettar1989phase,Moreo1991}. As a result, in non-half-filling scenarios, regime II [cf.~\cref{ld2_reg2}] with the dynamical transition from the SC state to the CDW state vanishes from the phase diagram [cf.~\cref{fig_mid_cur}(b)]. In other regimes, the dynamical properties are qualitatively the same as the half-filling case, with only quantitative changes in order parameters within each phase. Consequently, in non-half-filling cases, the transport phase diagram is expected to comprise only regimes I, III, and IV.

\section{Experimental considerations for ultracold atom platforms}
\label{sec_imple}

In addition to solid-state systems, ultracold atoms provide platform for simulating Kondo-SC
models, offering the unique advantage of studying long-time evolution and spatially-resolved correlations~\cite{Bloch2008a}. In this section, we discuss the possibility of realizing the physics described in this work using ultracold atoms.

The first task is to realize a Kondo impurity coupled to an attractive Hubbard bath [cf.~\cref{ham_gen}]. Here, the Kondo spin-exchange interaction can be realized in gases of alkaline-earth atoms~\cite{zhang2020controlling}, with the prominent example being fermionic $^{173}\rm Yb$ gas~\cite{scazza2014observation,cappellini2014direct,Riegger2018}. 

Using a state-dependent optical lattice, atoms in state $|g\rangle$ can experience shallow lattice potentials and act as host particles, while an atom in state $|e\rangle$, deeply localized in the lattice potential, acts as the impurity. The Kondo interaction arises from second-order processes~\cite{Riegger2018,zhang2020controlling}. By tuning lattice potentials and adding static or driven Zeeman fields~\cite{Kanasz-Nagy2018}, the anisotropy of the Kondo interaction can be controlled over a wide range. Specifically, the point contact between a Kondo impurity and two 1D leads [cf.~\cref{fig_mid_cur}(a)] can be realized using additional laser beams directed around the impurity region~\cite{husmann2015connecting}. The on-site Hubbard interaction naturally exists in various kinds of ultracold Fermi gases within optical lattices. In particular, for $^{173}\rm Yb$ atoms, the SU(N) Fermi-Hubbard model has been realized with various $N$~\cite{hofrichter2016direct}, where $N\le 6$. Using a magnetic field~\cite{zhang2020controlling}, the Hubbard interaction can be tuned to be attractive, and its strength can be adjusted~\cite{Mitra2018a,hartke2022quantum,hartke2023direct}. Moreover, various geometries of the bath can be efficiently realized using different optical lattice potentials~\cite{Bloch2008a}. Kondo type quantum impurity models can also be realized in systems of ultracold atoms using local pair-losses~\cite{stefanini2024dissipative,qu2024variational}, that can be realized experimentally~\cite{konishi2021universal,huang2023superfluid}.

To study the competition between SC order and Kondo correlation [cf.~\cref{fig1}], one can measure the SC and CDW order parameters, magnetization, and spin correlations in experiments.
The formation of the Kondo singlet leads to vanishing impurity magnetization and a characteristic magnetic susceptibility of the impurity~\cite{Kanasz-Nagy2018}, detectable using a spin-resolved quantum gas microscope~\cite{hilker2017revealing} and a Ramsey protocol~\cite{knap2013probing}. Moreover, the same quantum gas microscope allows for the direct measurement of spin and charge distributions in the bath, in both the ground state and real-time dynamics~\cite{hilker2017revealing,Vijayan2020}, enabling the determination of magnetization, CDW, and impurity spin-bath correlations~\cite{Kanasz-Nagy2018}. Additionally, matterwave optics~\cite{murthy2019direct} can be employed to measure the complex superfluid phase, facilitating the detection of the $\pi$ phase shift predicted in \cref{fig1}(a). Furthermore, by measuring the time-dependent density profile of the bath~\cite{luick2020ideal}, one can determine the current during evolution and experimentally probe different transport regimes [cf.~\cref{fig_mid_cur}(b)].

Altogether, key ingredients for detecting the physics of a Kondo impurity in an attractive Hubbard bath have been individually realized. However, considerable theoretical and experimental efforts are still needed to integrate these components into a single experimental setup. For example, the effective model for realizing the Kondo interaction~\cite{Kanasz-Nagy2018} with $^{173}\rm Yb$ atoms will be modified by the attractive Hubbard interaction in the bath, which should be accounted for in theoretical descriptions. In this context, our work opens up numerous opportunities to study the experimental realization of a Kondo impurity in an attractive Hubbard model.

Finally, other potential platforms exist to realize the physics studied here. One alternative approach is an analog simulation based on circuit QED platforms, which can emulate both the Kondo model~\cite{garcia2008quantum} and the Fermi-Hubbard model~\cite{reiner2016emulating}. Additionally, it is potentially possible to digitally simulate this model by integrating the Kondo interaction term into the existing digital simulation scheme of the Fermi-Hubbard model on a quantum processor~\cite{arute2020observation}.

\section{Discussion and Outlook}
\label{clu}

As we have summarized our main finding in \cref{sec_sum}, here we discuss interesting questions raised by our study and potential directions in the future.

First, by studying the ground states in the even and odd sectors of $\hat{N}$, we have numerically demonstrated the celebrated singlet-doublet phase transition~\cite{Satori1992,Sakai1993}, serving as a benchmark for our theoretical framework. We also point out that, within the same sector increasing the Kondo interaction and decreasing the attractive Hubbard interaction lead to a smooth crossover from doublet states to the singlet state. This allows for direct observation of the competition between SC order and Kondo correlations without requiring experimentally demanding schemes to modify the system parity.

Second, the directional emission of excitation in structured baths has been studied previously for the case of non-interacting impurities and baths~\cite{gonzaleztudela17a,gonzaleztudela17b,windt2024fermionic}, 
and the directional emission of PSP in the 2D relaxation dynamics observed in \cref{sec_2d_relax} represents the first demonstration of this phenomenon in an interacting fermionic system. Similar to the case of non-interacting impurities and baths, we expect that the directionality of the PSP emission is tunable (e.g., by changing the filling factor of the bath). Moreover, the intertwined orders enabled by the attractive Hubbard interaction can lead to phenomena different from those studied in the non-interacting baths~\cite{gonzaleztudela17a,gonzaleztudela17b,windt2024fermionic}, with the exponential decay of the PSP amplitude [cf.~\cref{fig2}(c)] being one such example.

Third, the rich phenomena we discovered in transport dynamics open up many exciting possibilities for further research. Here are a few examples: In the doublet phase, one can (a) study transport under an AC bias voltage, allowing the exploration of Shapiro steps~\cite{shapiro1963josephson} modified by the Kondo interaction; (b) investigate the parameter-dependent $U(1)$-symmetry restoration (corresponding to the dynamical transition from SC to CDW order) in the doublet phase using quantities such as the entanglement asymmetry~\cite{ares2023entanglement}, and explore the possible (inverse) quantum Mpemba effect~\cite{ares2023entanglement,joshi2024observing,aharony2024inverse} in the attractive Fermi-Hubbard model. In the singlet phase, the mechanisms we proposed for the anomalous DC conductance $G(V\ll 1) > 2e^2/h$ and the suppression of the AC component do not necessarily depend on the Kondo effect. Therefore, (c) we anticipate these phenomena in transport dynamics between two superconductors coherently coupled by any link with high transmission, such as the quantum point contact realized in ultracold atom platforms~\cite{huang2023superfluid,husmann2015connecting,fabritius2024irreversible}.

Fourth, our approach that applies the generalized Wick’s theorem to describe the 1D attractive Hubbard model is well-suited for describing experimentally relevant quasi-1D systems that exhibit spontaneous $U(1)$ symmetry breaking, such as superconducting nanowires and ultracold atomic gases~\cite{nazarov2009quantum,Franceschi2010,husmann2015connecting,prada2020andreev}. It is further interesting to study strictly 1D attractive Hubbard baths that exhibit Luttinger liquid behavior~\cite{essler2005one}, where the superfluid pair correlations decay as a power law. We expect the qualitative physical properties observed for the 1D baths in this work (cf.~\cref{sec_sg_12,ld2_sec}) to hold. For instance, the novel $\pi$-phase shift of the superconducting order parameter when crossing the Kondo singlet will correspond to a $\pi$-phase shift in the superfluid pair correlations in the same parameter regime for a strictly 1D bath, as has been numerically verified using DMRG~\cite{kscdmrg}.

Finally, beyond the questions mentioned above, our work can be extended in many directions. For instance, by tuning the lattice filling, the strength and range of bath electron interactions~\cite{essler2005one}, the anisotropy, and the number of orbits of the Kondo interaction, a wide range of equilibrium and non-equilibrium phenomena can be explored. Furthermore, by introducing additional impurities, the multi-Kondo problem in an attractive Hubbard bath can be investigated, which is expected to exhibit intricate competition between the Kondo effect, Ruderman-Kittel-Kasuya-Yosida (RKKY) interaction, SC, and CDW orders~\cite{yao2014phase,yao2014enhanced,PhysRevResearch.6.033022}. Additionally, the non-Gaussian variational approach can be extended to study finite-temperature behaviors of such systems~\cite{Shi2020}, enabling the study of thermal transport properties~\cite{cahill2003nanoscale} and characteristic quantities such as the specific heat and the Wilson ratio~\cite{hewson1997kondo}.

\emph{Note added---}
Upon completion of the present manuscript, we became aware of an independent work appearing on arXiv~\cite{moca2024spectral}, which investigates the ground-state and finite-temperature spectral properties of fractionalized Shiba states induced by a Kondo impurity coupled to a one-dimensional attractive Hubbard bath. This work complements our study, which focuses on the long-term quantum dynamics of a Kondo impurity coupled to one and two-dimensional attractive Hubbard baths.

\begin{acknowledgments}
We thank Miguel Cazalila, Jan von Delft, Tilman Esslinger, Chen-How Huang, Gergely Zarand, and Zhongda Zeng for their insightful discussions. The research is part of the Munich Quantum Valley, which is supported by the Bavarian state government with funds from the Hightech Agenda Bayern Plus. ZYW and JIC acknowledge funding from the German Federal Ministry of Education and Research (BMBF) through EQUAHUMO (Grant No. 13N16066) within the funding program Quantum Technologies—From Basic Research to Market. ZYW also acknowledge DoE ASCR Accelerated Research in Quantum Computing program (awards No.~DE-SC0020312 and No.~DE-SC0025341). T.S. is supported by NSFC (Grants No. 12135018, 12047503, and 12274331), by National Key Research and Development Program of China (Grant No. 2021YFA0718304), and by CAS Project for Young Scientists in Basic Research (Grant No. YSBR-057). ED acknowledges support from the ETH grant, the SNSF project 200021\_212899, the Swiss State Secretariat for Education, Research and Innovation (contract number UeM019-1), and NCCR SPIN, a National Centre of Competence in Research, funded by the Swiss National Science Foundation (grant number 225153). 

\end{acknowledgments}

\bibliography{library.bib}

\appendix

\renewcommand{\thefigure}{A\arabic{figure}}

\setcounter{figure}{0}

\section{SU(2) pseudo-spin symmetry of the system}
\label{sc_cdw_trans}

The attractive Hubbard model at half-filling exhibits a global SU(2) pseudo-spin symmetry~\cite{yang1990so,zhang1990pseudospin}, resulting in degenerate ground states connected by pseudo-spin rotations. Here, we demonstrate that for the lattice hopping and Kondo couplings considered in this paper [cf.~\cref{para_1d,para_2d,ham_2lead}], and with no bias voltage ($V=0$), the system Hamiltonian $H$ [\cref{ham_gen}] at half-filling retains the same global SU(2) pseudo-spin symmetry.

The global pseudo-spin operators are defined as~\cite{yang1990so,zhang1990pseudospin}
\begin{align}
\hat T^{z} &=\sum_{\bm j} \hat T_{\bm j}^{z}, \\
\hat T^{\pm} &=\sum_{\bm j} \hat T_{\bm j}^{\pm},
\end{align}
where
\begin{align} \label{pso}
\hat T_{\bm j}^{+} &=\lambda_{\bm j} c_{\bm j ,\uparrow}^{\dagger} c_{\bm j ,\downarrow}^{\dagger}, \\
\hat T_{\bm j}^{-} &=\lambda_{\bm j} c_{\bm j ,\downarrow} c_{\bm j ,\uparrow}, \\
\hat T_{\bm j}^{z} &=\frac{1}{2}\left(c_{\bm j ,\uparrow}^{\dagger} c_{\bm j ,\uparrow}+c_{\bm j ,\downarrow}^{\dagger} c_{\bm j ,\downarrow}-1\right),
\end{align}
with $\lambda_{\bm j}=(-1)^j$ for a single 1D bath [cf.~\cref{para_1d}] and the two-lead case [cf.~\cref{para_2d}], and $\lambda_{\bm j}=(-1)^{j_x+j_y}$ for a single 2D bath [cf.~\cref{ham_2lead}]. These operators obey the commutation relations $[\hat T^{z}, \hat T^{\pm} ] = \hat T^{\pm}$ and $[\hat T^{+}, \hat T^{-} ] = 2 \hat T^{z}$, forming an SU(2) algebra.

The system Hamiltonian $H$ consists of the attractive Hubbard bath Hamiltonian $H_{\rm bath}$ [\cref{ham_bath}] and the Kondo coupling $H_{\rm int}$ [\cref{ham_kondo}]. For $H_{\rm bath}$, its pseudo-spin SU(2) symmetry implies that, at half-filling, $[\hat T^z,H_{\rm bath}]= [\hat T^{\pm},H_{\rm bath}] = 0$. Moreover, $H_{\rm int}$ also commutes with $\hat T^z$ and $\hat T^\pm$. In the case where the impurity couples to a single site [cf.~\cref{para_1d,para_2d}], we define $\hat S_{\bm 0}^\gamma= \sum\limits _{\alpha,\beta \in (\uparrow, \downarrow)}c_{{\bm 0},\alpha}^{\dagger}\sigma_{\alpha\beta}^{\gamma}c_{{\bm 0}, \beta}/2$ as the spin operator of the fermion on the lattice site $\bm 0$. Thus,
\begin{align} \label{}
	[\hat T^z,H_{\rm int}] = J  \sum_{\gamma \in (x,y,z)}\hat s^\gamma_{\rm imp }[\hat T^z_{\bm 0}, \hat S^\gamma _0] = 0,
\end{align}
since $[\hat T^z_{\bm 0}, \hat S^\gamma _0] = 0$ for all $\gamma$. Similarly, $[\hat T^\pm,H_{\rm int}] = 0$ since $[\hat T^{\pm}_{\bm 0}, \hat S^\gamma _0] = 0$ for all $\gamma$. For the two-lead case [cf.~\cref{ham_2lead}], the same calculation shows that $H_{\rm int}$ commutes with $\hat T^z$ and $\hat T^\pm$. Therefore, for the setups considered in this paper, the system Hamiltonian $H$ commutes with $\hat T^z$ and $\hat T^\pm$ at half-filling and without a bias voltage, thus possessing global SU(2) pseudo-spin symmetry. This implies that, for the half-filling case, it suffices to study the ground state properties of $H$ with the bath having SC order.

\section{Details of the variational method}
\label{mod_details}

In this section, we provide additional details of the variational approach~\cite{kraus2010generalized,Shi2018a} utilized in this study [cf.~\cref{var_method_main}].

\subsection{Variational Time-Evolution Equations}
\label{apd_var_eq}

First, we define some notations to simplify the descriptions in this section. The expectation value of an operator $O$ with respect to various ansatz states is defined as
\begin{align}\label{}
\langle	\Psi_{\rm NGS} | &O |	\Psi_{\rm NGS} \rangle \equiv \langle O \rangle_{\rm NGS}, \\
	\quad \langle	\Psi_{\rm GS} |  &O   |	\Psi_{\rm GS} \rangle \equiv \langle O \rangle_{\rm GS},\nonumber \\
	\quad \langle	\Psi_{\rm GS} | ( _{\rm imp}\langle  + |   &O  |+ \rangle _{\rm imp} ) |	\Psi_{\rm GS} \rangle \equiv \langle O \rangle^{\rm imp}_{\rm GS}.\nonumber
\end{align}
From the relation between $|	\Psi_{\rm NGS} \rangle$ and $|	\Psi_{\rm GS} \rangle$ [\cref{ngs_state}], we obtain
\begin{equation} \label{}
\langle O \rangle_{\rm NGS} =  \langle  \hat U_{\rm NGS}^{\dag} O \hat U_{\rm NGS}  \rangle^{\rm imp}_{\rm GS}.
\end{equation}

The parameters in our variational ansatz $|	\Psi_{\rm NGS}(\Gamma_f) \rangle$ [\cref{ngs_state}] are represented by the covariance matrix ${\Gamma_f}$ of the Gaussian state (size $4N \times 4N$), which is defined as~\cite{kraus2010generalized,weedbrook2012}
\begin{equation} \label{covmat}
{\Gamma_f} = {\langle {CC^\dag } \rangle _{\rm GS}} = \left( {\begin{array}{*{20}{c}}
{\Gamma_f^{11}}&{\Gamma_f^{12}}\\
{\Gamma_f^{21}}&{\Gamma_f^{22}}
\end{array}} \right),
\end{equation}
where the Nambu spinor $C$ is defined in the main text [near \cref{ngs_state}].

The order parameters of the system can be expressed in terms of $\Gamma_f$ as follows:
\begin{equation} \label{ord_covmat}
\begin{array}{l}
{\Delta _ {\bm j}} \equiv \langle {{c_{\bm j ,\downarrow }}{c_{\bm j ,\uparrow }}} \rangle_{\rm GS} = {( {\Gamma_f^{12}} )_{\bm j ,\downarrow ,\bm j ,\uparrow }},\\
{n_{\bm j, \alpha }} \equiv \langle {c_{\bm j, \alpha }^\dag {c_{\bm j, \alpha }}} \rangle _{\rm GS}= {( {\Gamma_f^{22}} )_{\bm j, \alpha ,\bm j, \alpha }},\\
{f_{\bm j}} \equiv \langle {c_{\bm j ,\uparrow }^\dag {c_{\bm j ,\downarrow }}} \rangle _{\rm GS} = {( {\Gamma_f^{22}} )_{\bm j ,\uparrow ,\bm j ,\downarrow }},\\
\Gamma_f^{\rm P} \equiv {\langle {{{\hat P}_{{\rm{bath}}}}{C^{\dag}}C} \rangle _{\rm {GS}}}.
	\end{array}
\end{equation}
Thus, one can write the variational energy of the Hamiltonian $H$ [\cref{ham_gen}] as
\begin{equation} \label{Evar}
\begin{array}{*{20}{l}}
{{E_{{\rm{var}}}} \equiv \langle H \rangle _ {\rm NGS} = \langle  H' \rangle }^{\rm imp}_{\rm GS} \\ 
{=\sum\limits_{\bm j \bm l,\alpha } {{h_{\bm j \bm l}}} {{( {\Gamma_f^{22}} )}_{\bm j,\alpha ,\bm l\alpha }} + \frac{1}{4} \sum\limits_{\bm j \bm l,\alpha \beta } J_{\bm j \bm l}^x {\hat \sigma _{\alpha \beta }^x{{( {\Gamma_f^{22}} )}_{\bm j, \alpha ,\bm j\beta }}} }\\
{ + \frac{1}{4} \sum\limits_{\bm j \bm l,\alpha \beta } {{{( { - i J_{\bm j \bm l}^y {\hat \sigma ^y} + J_{\bm j \bm l}^z{\hat \sigma ^z}} )}_{\alpha \beta }}} {{( {\Gamma_f^{\rm{P}}} )}_{\bm j,\alpha ,\bm l\beta }}}\\
{ - U \sum\limits_{\bm j} {( {\Delta _{\bm j}^*{\Delta _{\bm j}} + {n_{{\bm j}, \uparrow }}{n_{{\bm j}, \downarrow }} - {f_{\bm j}}f_{\bm j}^*} )} },
\end{array}
\end{equation}
where we have used the transformed Hamiltonian $H'$ [\cref{ham_trans}] and chosen $s_{\rm imp}^x=1/2$. Moreover, as mentioned in \cref{main_wick}, we use Wick's theorem to compute the expectation value of $c_{\bm j ,\uparrow}^{\dagger} c_{\bm j ,\downarrow}^{\dagger} c_{\bm j ,\downarrow} c_{\bm j ,\uparrow}$, which later lead to quadratic terms in the mean-field Hamiltonian ${\cal H}$ (defined as ${\cal H} = 2 \delta E_{\rm var} / \delta \Gamma_f$).

We compute the functional derivative of $E_{\rm var}$ with respect to the covariance matrix $\Gamma_f$ in the same way as in Refs.~\cite{Shi2018a,ashida2018variational}. The resulting ${\cal H}$ can then be used to perform the imaginary (real) time evolution of $\Gamma_f$ using the equation of motion \cref{ite_eom} (\cref{rte_eom}).

To study the ground state properties [cf.~\cref{sec_sg_12}], such as the singlet-doublet phase transition, we need to probe the ground states with different total spins ($\langle \hat S^2 \rangle=3/4$ for the doublet phase and $\langle \hat S^2 \rangle=0$ for the singlet phase) while maintaining the same mean particle number $\langle \hat N \rangle = N$ (half-filling). Thus, during the imaginary time evolution, we enforce the conservation of total spin ${{\hat S}^2}$ and the mean particle number $\hat N$. Moreover, the real-time dynamics studied in \cref{sec_2d_relax,ld2_sec} also conserve the total spin and mean particle number. We introduce a penalty term in the Hamiltonian to ensure these conservation laws during the real-time dynamics. These aspects are elaborated on in the following sections.

\subsection{Imaginary time evolution with conserved quantities}

\begin{figure*}
	\centering
	\includegraphics[width=0.9\textwidth]{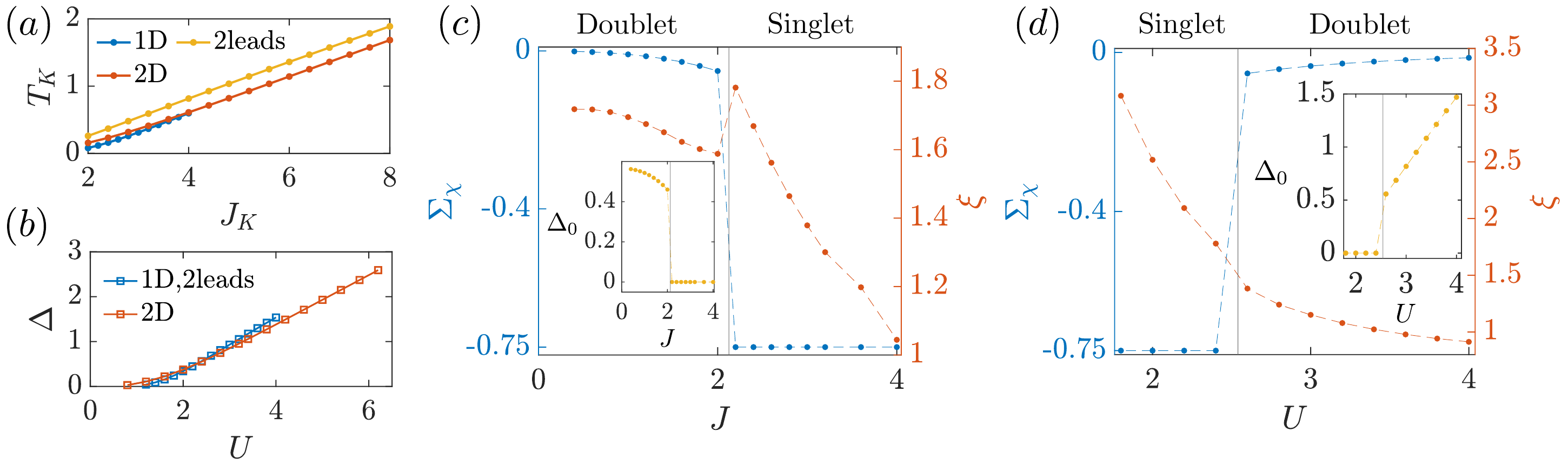}
        \caption{(a) The Kondo temperature $T_K$ as a function of the effective Kondo coupling $J_K$ for the bath geometries considered in this work.  (b) The SC gap $\Delta$ as a function of the attractive Hubbard interaction $U$ for the bath geometries considered in this work. (c) The total impurity-bath spin correlation $\Sigma_{\chi}$ and the size $\xi$ of the (partial) Kondo cloud as functions of the Kondo coupling $J$, with a fixed attractive Hubbard interaction $U=2.4$, for the case where the impurity is coupled to the 1D bath [cf.~\cref{para_1d}]. The inset shows the SC order parameter $\Delta_{0}$ at the impurity site. The vertical line indicates the singlet-doublet phase transition point. The system size $N=2L+1=401$. (d) The same as in (c), but as a function of $U$, with fixed $J=2.2$.}
        \label{apd_fig1}
\end{figure*}

\subsubsection{General formalism}
\label{cons_gf}

Consider multiple conserved quantities, ${O_1},...,{O_n}$, that commute with the Hamiltonian $H$, i.e., $[O_j,H]=0$ for all $j$. To enforce their conservation during imaginary time evolution, one can introduce time-dependent Lagrange multipliers~\cite{Shi2020}, thereby modifying the Hamiltonian to
\begin{equation} \label{cons_ham}
{H}_c^{\rm ite} = {H} - \sum_{j=1}^n \mu_{O_j}{O_j},
\end{equation}
where $\{\mu_{O_j}\}$ are the chemical potentials. The imaginary time evolution under this modified Hamiltonian $H_c^{\rm ite}$ is given by
\begin{equation} \label{ite_con}
{d_\tau}|\Psi \rangle = - ({H}_c^{\rm ite} - \langle {H}_c^{\rm ite}\rangle )|\Psi \rangle.
\end{equation}
Our goal is to choose appropriate $\{\mu_{O_j}\}$ so that, during the imaginary time evolution,
\begin{equation} \label{cons_oj}
{d_\tau}\langle {O_j} \rangle = 0,\quad \forall j = 1,...,n.
\end{equation}

In the case of a single conserved quantity $O_1$, by choosing~\cite{Shi2020}
\begin{equation} \label{}
{\mu_{O_1}} = \frac{\langle {O_1}H\rangle - \langle {O_1}\rangle \langle H\rangle}{{\langle {O_1^2} \rangle - {\langle {O_1}\rangle}^2}},
\end{equation}
it can be shown that ${d_\tau}\langle {O_1} \rangle = 0$. For multiple conserved quantities ${O_1},...,{O_n}$, the conditions in \cref{cons_oj} provide $n$ equations, allowing for the determination of the time-dependent chemical potentials ${\left\{ {{\mu _{O_j}}(\tau)} \right\}_{j = 1,...,n}}$.

\subsubsection{Imposing the conservation of $\hat S^2$ and $\hat N$}
\label{apd_conserv_ite_sn}

We now apply the approach in \cref{cons_gf} to the variational imaginary-time evolution method using the non-Gaussian ansatz [\cref{apd_var_eq}]. In this context, incorporating the Lagrange multipliers [\cref{cons_ham}] modifies the variational energy $E_{\rm var}$ [\cref{Evar}] to
\begin{equation} \label{Evar_c}
E_{\rm var}^c = E_{\rm var} - \mu_{\hat S^2} \langle \hat S^2 \rangle_{\rm NGS} - \mu_{\hat N} \langle \hat N \rangle_{\rm NGS}.
\end{equation}
To evaluate the last two terms, we first recall the definitions of the total spin operator ${{\hat S}^2}$ and the mean particle number operator $\hat N$:
\begin{equation} \label{sn_form}
\begin{array}{l}
{{\hat S}^2} = \hat S_x^2 + \hat S_y^2 + \hat S_z^2  \\
= \sum\limits_{\gamma \in (x,y,z)}  {{{( {\hat \sigma _{{\rm{imp}}}^\gamma  + \sum\limits_{ {\bm j},\alpha \beta \in (\uparrow, \downarrow) } {c_{{\bm j},\alpha }^{\dagger}\sigma _{\alpha \beta }^\gamma {c_{{\bm j},\beta }}} } )}^2}} ,\\
\hat N = \sum\limits_{{\bm j},\alpha \in (\uparrow, \downarrow) } {c_{{\bm j},\alpha }^\dag {c_{{\bm j},\alpha }}}.
\end{array}
\end{equation}

Using the unitary transformation $\hat U_{\rm NGS}$ [\cref{transU}], we can compute $\langle \hat S^2 \rangle _{\rm NGS} = \langle \hat S'^2 \rangle^{\rm imp} _{\rm GS}$ and $\langle \hat N \rangle _{\rm NGS} = \langle \hat N' \rangle^{\rm imp} _{\rm GS}$ using ${{\hat S}^2}$ and $\hat N$ in the transformed frame, defined as
\begin{equation} \label{ns_trans}
\begin{array}{l}
\hat N' = {\hat U_{\rm NGS}^\dag }\hat N \hat U_{\rm NGS} = \hat N,\\
{{\hat S'_x} =  - \hat \sigma _{{\rm{imp}}}^z{\hat P_{{\rm{bath}}}} - i\hat \sigma _{{\rm{imp}}}^y{\hat P_{{\rm{bath}}}}{C^\dag }{\sigma ^x}C,}\\
{{{\hat S}'_y} = \hat \sigma _{{\rm{imp}}}^y - i\hat \sigma _{{\rm{imp}}}^y{{\hat P}_{{\rm{bath}}}}{C^\dag }{\sigma ^y}C,}\\	
{{\hat  S'_z} = \sigma _{{\rm{imp}}}^x{\hat P_{{\rm{bath}}}} + {C^\dag }{\sigma ^z}C,}
\end{array}	
\end{equation}
where the bath parity operator ${\hat P_{{\rm{bath}}}}$ [\cref{pb}] and the Nambu spinor $C$ [\cref{ngs_state}] are defined previously. We can then express $\langle \hat S'^2\rangle^{\rm imp}_{\rm GS}$ and $\langle \hat N' \rangle^{\rm imp}_{\rm GS}$ in terms of the covariance matrix $\Gamma_f$ [\cref{covmat}] as
\begin{equation} \label{s2_cov}
\begin{aligned}
&\langle \hat S'^2\rangle^{\rm imp}_{\rm GS} = 3-2  {\rm Tr}[ \tau_x  \Gamma_f^{11}]+
       2{\rm Tr}[(\tau_z - i \tau_y) \Gamma_f^{\rm P}] \\
       & +{\rm Tr}[ \tau_x  (\Gamma_f^{12})^{\dag}   \tau_x  \Gamma_f^{12}]-{\rm Tr}[ \tau_y  (\Gamma_f^{12})^{\dag}   \tau_y  \Gamma_f^{12}]+{\rm Tr}[ \tau_z  (\Gamma_f^{12})^{\dag}   \tau_z  \Gamma_f^{12}]\\
       &+\sum_{\gamma \in (x,y,z)} \left({\rm Tr}[ \tau_\gamma  \Gamma_f^{11}]\right)^2+
       {\rm Tr}[ \tau_\gamma  \Gamma_f^{11}   \tau_\gamma  ({\mathbb I}_N - \Gamma_f^{11})],
\end{aligned}	
\end{equation}
\begin{equation} \label{n_cov}
	\langle N' \rangle^{\rm imp}_{\rm GS}  =  {\rm Tr}[\Gamma_f^{11}],
\end{equation}
where
\begin{equation} \label{}
\tau_x \equiv \sigma_x \otimes {\mathbb I}_N, \quad \tau_y \equiv \sigma_y \otimes {\mathbb I}_N, \quad \tau_z \equiv \sigma_z \otimes {\mathbb I}_N,
\end{equation}
and $\Gamma_f^{\rm P}$ is defined previously in \cref{ord_covmat}.

Using \cref{n_cov,s2_cov} and the techniques from Refs.~\cite{Shi2018a,ashida2018variational}, one can obtain the functional derivatives
\begin{equation} \label{hshn_mf}
			{{\cal H}_{{{\hat S}'^2}}} = \frac{2\delta \langle {{{\hat S}'^2}} \rangle^{\rm imp}_{\rm GS} }{\delta {\Gamma_f}}, \quad {{\cal H}_{{{\hat N}'}}} = \frac{2\delta \langle {{{\hat N'}}} \rangle^{\rm imp}_{\rm GS} }{\delta {\Gamma_f}},
\end{equation}
and the equation of motion for the covariance matrix [\cref{ite_eom}] is modified as~\cite{Shi2020}
\begin{equation} \label{eom_cons}
{d _\tau }{\Gamma_f} = \{ {{\cal H}_c^{\rm ite},{\Gamma_f}} \} - 2{\Gamma_f}{\cal H}_{c}^{\rm ite}{\Gamma_f},
\end{equation}
with ${\cal H}_c^{\rm ite} = {\cal H} - \mu_{\hat S^2} {\cal H}_{\hat S'^2} - \mu_{\hat N} {\cal H}_{\hat N'}$.

The conservation equations in this case are
\begin{equation} \label{cons_sn}
{d _\tau }\langle {{\hat S'^2}} \rangle^{\rm imp} _{\rm GS}= 0, \quad {d _\tau }\langle \hat N' \rangle^{\rm imp} _{\rm GS} = 0.
\end{equation}
By expressing $\hat N'$ and $\hat S'^2$ (excluding the impurity part) in the Nambu spinor basis as
\begin{align}\label{}
	_{\rm imp}\langle  + |   \hat S'^2  |+ \rangle _{\rm imp}   = \frac{1}{2} C^{\dagger} {{\cal S}'^2} C, \\ 
_{\rm imp}\langle  + |   \hat N'  |+ \rangle _{\rm imp} = \frac{1}{2} C^{\dagger} {{\cal N}'} C,\nonumber
\end{align}
we can express \cref{cons_sn} in terms of the covariance matrix $\Gamma_f$ and the operators ${{\cal S}'^2}$ and ${{\cal N}'}$ as
\begin{equation} \label{cov_cons}
{\rm Tr}\left[ {{\cal N}'{d _\tau }{\Gamma_f}} \right] = 0,\qquad {\rm Tr}\left[ {{{\cal S}'^2}{d _\tau }{\Gamma_f}} \right] = 0.
\end{equation}
By initializing the state with total spin $\langle \hat S^2 \rangle_{\rm NGS}=3/4$ ($\langle \hat S^2 \rangle_{\rm NGS}=0$) and half-filling $\langle \hat N \rangle _{\rm NGS} = N$, the imaginary time evolution using \cref{eom_cons} [with ${\mu _{S^2}}(\tau)$ and ${\mu _N}(\tau)$ determined from \cref{cov_cons}] conserves $\langle \hat S^2 \rangle_{\rm NGS}$ and $\langle \hat N \rangle_{\rm NGS}$, thereby allowing one to find the corresponding ground state in the doublet (singlet) phase.

\subsection{Real-time evolution with conserved quantities}

To enhance the numerical stability of solving \cref{rte_eom} for the real-time dynamics, we employ a technique similar to that used in Ref.~\cite{shi2020ultrafast}, introducing a penalty term $H_{\Lambda} = \Lambda \hat S^2$, where $\Lambda$ is chosen to be much larger than all energy scales in the system. Notably, the specific value of $\Lambda$ is unimportant for the results discussed in this paper. The mean-field Hamiltonian for the penalty $H_\Lambda$ in the transformed frame is given by ${\cal H}_{\Lambda} = \Lambda {{\cal H}_{{{\hat S}'^2}}}$ [cf.~\cref{hshn_mf}]. Therefore, in our numerical calculations of the real-time dynamics, the mean-field Hamiltonian in \cref{rte_eom} is modified to ${\cal H} \rightarrow {\cal H} + {\cal H}_{\Lambda}$, ensuring good conservation of the total spin during the real-time evolution.

\section{Additional results on the ground state properties}
\label{apd_1d}

\begin{figure*}
	\centering
	\includegraphics[width=1\textwidth]{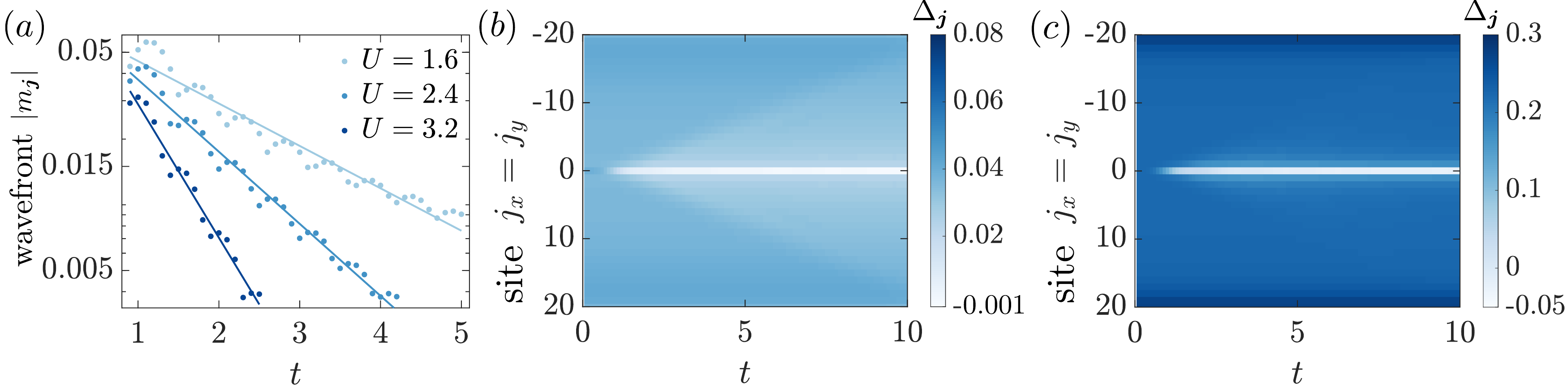}
        \caption{Additional results for the 2D quench dynamics [cf.~\cref{sec_2d_relax}], for a fixed Kondo coupling of $J=4.0$. (a) The evolution of the PSP wavefront amplitude $|m_{\bm j}|$ [cf.~\cref{fig2}(c)], for various attractive Hubbard interactions $U$. (b) The evolution of the SC order parameter $\Delta_{\bm j}$ along the diagonal sites of the 2D bath, for $U=0.8$. (c) Same as (b), but with $U=1.6$.}
        \label{apd_fig2}
\end{figure*}

In the main text, we characterize the system using the Kondo coupling $J$ and the Hubbard interaction $U$ instead of the Kondo temperature $T_K$ and the SC gap $\Delta$. This choice is made because the attractive Hubbard bath cannot simply be viewed as a superconductor, as the gap $\Delta$ does not unambiguously characterize the bath. For the reader's convenience, \cref{apd_fig1}(a,b) displays $T_K$ and $\Delta$ as functions of $J$ and $U$. Here, the gap $\Delta$ is directly extracted from the bulk value of the SC order parameter in our calculations. To determine $T_K$, we compute the local impurity magnetization $m=\langle {\hat s^{z} _{\rm imp}} \rangle$ as a function of an externally applied local magnetic field $h_z$ on the impurity (see calculation details in Ref.~\cite{ashida2018variational}) within the Kondo model ($U=0$) for various geometries studied in the main text [1D: \cref{para_1d}, 2D: \cref{para_2d}, 2leads: \cref{ham_2lead}]. We then obtain $T_K$ using the formula~\cite{hewson1997kondo}:
\begin{equation} \label{}
	T_{\mathrm{K}}  = \left(\frac{1}{4}\frac{1}{\partial m / \partial h_z}\right )_{h_z\rightarrow 0}.
\end{equation}
Due to the different coupling geometries studied in the main text, \cref{apd_fig1}(a) plots $T_K$ as a function of the effective Kondo coupling $J_K$, which is defined as:
\begin{equation} \label{}
	J_{K}=\left\{ \begin{aligned}
&J, \qquad \textrm{for 1D,2D bath}\\
&2J, \qquad \textrm{for 2leads}
\end{aligned}\right.
\end{equation}

Additionally, in \cref{sec_gs_order}, we focused on the behavior of the order parameters [cf.~\cref{fig1}(c,d)] in the ground state when the impurity couples to a 2D bath [cf.~\cref{para_2d}]. Figure \ref{apd_fig1}(c,d) shows the behavior of the same set of order parameters when the impurity couples to a 1D bath [cf.~\cref{para_1d}]. Figure \ref{apd_fig1}(c) shows the size $\xi$ of the (partial) Kondo cloud [cf.~\cref{ck_var}], the SC order parameter at the impurity site $\Delta_{0}$, and the total impurity-bath correlations $\Sigma_\chi$ [cf.~\cref{sum_chi}] as functions of the Kondo coupling $J$. Similar to the 2D case, the magnitude of $\Sigma_\chi$ gradually increases as $J$ increases and remains at $\Sigma_\chi=-0.75$ in the singlet phase. The (partial) Kondo cloud size $\xi$ decreases with $J$ in both the doublet and singlet phases, and $\Delta_{0}$ decreases with $J$ in the doublet phase and becomes $0$ in the singlet phase, as already observed in \cref{fig1}(a). In the singlet phase, the SC order parameter changes sign when crossing the impurity site $(j=0)$, so its amplitude at the impurity site must be $\Delta_{0}=0$ due to the mirror symmetry of the 1D bath of length $N=2L+1$ along the impurity site. \cref{apd_fig1}(d) shows the same set of order parameters as a function of $U$, where we again see similar behavior to that in 2D [cf.~\cref{fig1}(d)], with the only major difference being that $\Delta_{0}=0$ in the singlet phase, as we just explained.

The similarity in behaviors when the impurity couples to a 1D or 2D bath is expected, as the impurity physics generally does not strongly depend on the dimension of the bath to which it couples~\cite{Busser2013}. However, we still observe a non-trivial difference in the ground state between 1D and 2D due to the difference between the 1D and 2D attractive Hubbard baths. For example, the phase shift of the SC order parameter [cf.~\cref{fig1}(a,b)], as discussed in the main text.

\section{Additional results of the 2D quench dynamics}
\label{apd_2d_qu}

In the 2D quench dynamics discussed in \cref{sec_2d_relax}, we observe the ballistic and damped propagation of a pulse of spin polarization (PSP) [cf.~\cref{fig2}(c)]. The evolution of the wavefront amplitude is shown in \cref{apd_fig2}(a). In general, the wavefront amplitude decays exponentially as $\sim e^{-\gamma t}$, accompanied by small oscillations. As explained in the main text, this exponential decay results from the finite quasiparticle lifetime induced by the SC order in the bath. Increasing $U$ (and thus the magnitude of the SC order parameter) leads to an increased decay rate $\gamma$ [see also the inset of \cref{fig2}(c)].

We also present the evolution of the SC order parameter at the diagonal sites of the bath in \cref{apd_fig2}(b,c). Near the impurity site ${\bm 0} = (0,0)$, the SC order parameter is significantly suppressed due to the formation of the (partial) Kondo cloud, with its value even becoming negative at the impurity site. This change indicates that the system locally relaxes to its equilibrium (ground state) configuration [cf.~\cref{fig1}(b)]. The propagating PSP [cf.~\cref{fig2}(b,c)] suppresses the SC order parameter in the region it travels through. This suppression is more pronounced when the amplitude of the SC order parameter is smaller, illustrating the dynamic competition between magnetism and superconductivity in the bath.

\section{Additional results of two-lead transport dynamics}
\label{apd_2ld}

\subsection{The ground state of the two-lead case}
\label{apd_2l_gs}

In this section, we provide an analytical explanation for the $\pi$ phase shift of the SC order parameter observed in the doublet phase ground state when the impurity is coupled to two 1D leads [cf.~\cref{ld2_sc_ph_fig}]. For our purposes, it suffices to consider the regime $J \ll 1$, where perturbation theory is applicable.

In the two-lead case, electron tunneling between the two leads must occur via the impurity through the spin-exchange interaction. The relevant Hamiltonian term in the transformed frame is given by [cf.~\cref{ham_trans,ham_2lead}]
\begin{equation} \label{ham_2lead_k}
H'^{\rm 2leads}_{\rm int} =  \frac{1}{2} \left[{{\hat \Phi^x_{\rm 2leads} }} + {{\hat P}_{{\rm{bath}}}}\left( {{ - i{{\hat \Phi^y_{\rm 2leads} }}} + {{\hat \Phi^z_{\rm 2leads} }}} \right)\right],
\end{equation}
where
\begin{equation} \label{}
\Phi^\gamma_{\rm 2leads} = \frac{J}{2} \sum_{\eta, \eta^{\prime} \in (-,+), \alpha, \beta}  {c}_{\eta \cdot 1, \alpha }^{\dagger} \hat \sigma^{\gamma}_{\alpha \beta} {c}_{\eta' \cdot 1, \beta }, \quad \gamma \in (x,y,z),
\end{equation}
and we have taken $s_{{\rm{imp}}}^x=1/2$ as a classical number, since $\hat s_{{\rm{imp}}}^x$ commutes with the system Hamiltonian in the transformed frame.

For $J \ll 1$, we assume that in the transformed frame, the two leads are approximately in BCS states with a phase difference $\theta$,
\begin{equation} \label{}
	|\Psi_{\rm bath} \rangle \approx |\sqrt{L} \rangle _{-} |\sqrt{L} e^{i\theta}\rangle _{+},
\end{equation}
where $L$ is the length of each chain, and $|\sqrt{L} \rangle _{-}$ ($|\sqrt{L} e^{i \theta} \rangle _{+}$) denotes the BCS state, i.e., a coherent state of Cooper pairs, with phase 0 ($\theta$) on the left (right) bath. Furthermore, we take ${\hat P_{\rm bath}}\approx I$ in \cref{ham_2lead_k} since $\langle \hat P_{\rm bath} \rangle \approx 1$ when $J\ll 1$ in the doublet phase.

When removing or adding a Cooper pair to the system, the BCS state remains approximately the same, thus we have
	\begin{align}
	|\Psi_{\rm bath} \rangle &\approx | \Psi_{\rm bath}^L \rangle\propto c^{\dag}_{-1,\uparrow} c^{\dag}_{-1,\downarrow} |\sqrt{L} \rangle _{-} |\sqrt{L} e^{i\theta}\rangle _{+}\\
	&\approx | \Psi_{\rm bath}^R \rangle \propto e^{i\theta} c^{\dag}_{1,\uparrow} c^{\dag}_{1,\downarrow} |\sqrt{L} \rangle _{-} |\sqrt{L} e^{i\theta}\rangle _{+}.
\end{align}

For $|\Psi_{\rm bath} \rangle$ to be the ground state, it must be stable under perturbation. Therefore, we can determine $\theta$ by examining the energy correction due to the perturbation that couples $| \Psi_{\rm bath}^L \rangle$ to $| \Psi_{\rm bath}^R \rangle$, where both states have the same ground state energy $E_0$. This corresponds to a process in which a Cooper pair tunnels from the left to the right bath. The leading order of the energy correction arises from a second-order process, which is given by
\begin{equation} \label{w_ovlp}
\delta E =\sum_{\lambda} \frac{\langle \Psi_{\rm bath}^R | H'^{\rm 2leads}_{\rm int} | \lambda \rangle \langle \lambda | H'^{\rm 2leads}_{\rm int} |\Psi_{\rm bath}^L \rangle}{(E_0 - E_\lambda)(E_\lambda - E_0)},
\end{equation}
where \{$|\lambda \rangle$\} are the intermediate states with energy $E_\lambda$. There are four possible intermediate states $|\lambda \rangle$, given by
\begin{align} \label{int_state}
&|\uparrow,\downarrow  \rangle \propto c^{\dag}_{-1,\uparrow} c^{\dag}_{+1,\downarrow} | \Psi_{\rm bath} \rangle, \\
&|\downarrow,\uparrow  \rangle \propto c^{\dag}_{-1,\downarrow} c^{\dag}_{+1,\uparrow} | \Psi_{\rm bath} \rangle, \nonumber \\ 
&|\uparrow,\uparrow  \rangle \propto c^{\dag}_{-1,\uparrow} c^{\dag}_{+1,\uparrow} | \Psi_{\rm bath} \rangle, \nonumber \\
&|\downarrow,\downarrow  \rangle \propto c^{\dag}_{-1,\downarrow} c^{\dag}_{+1,\downarrow} | \Psi_{\rm bath} \rangle, \nonumber
\end{align}
with $E_\lambda - E_0 \approx 2\Delta$ (where $\Delta > 0$ is the SC gap). In these intermediate states, a Cooper pair is broken, with each bath containing one of its electrons. Note that the spin-exchange interaction \cref{ham_2lead_k} can produce intermediate electron states where the electron's spins are flipped.

Using \cref{ham_2lead_k,w_ovlp,int_state}, we find 
\begin{equation} \label{}
\delta E \approx {3e^{-i\theta}}J^2/32\Delta^2.	
\end{equation}
Therefore, only when $\theta = \pi$ does this second-order tunneling process lower the energy as $J$ increases. Thus, in the doublet phase of the two-lead bath case, the SC order parameter across the two leads exhibits a $\pi$ phase shift. In the singlet phase, however, the scattering of Cooper pair by the Kondo singlet introduces an additional $\pi$ phase shift, as explained in \cref{sec_gs_order}. Consequently, each Cooper pair experiences a 0 phase shift in the singlet phase for the two-lead case, consistent with observations in the literature~\cite{Franceschi2010}.

\begin{figure}
	\centering
\includegraphics[width=0.5\textwidth]{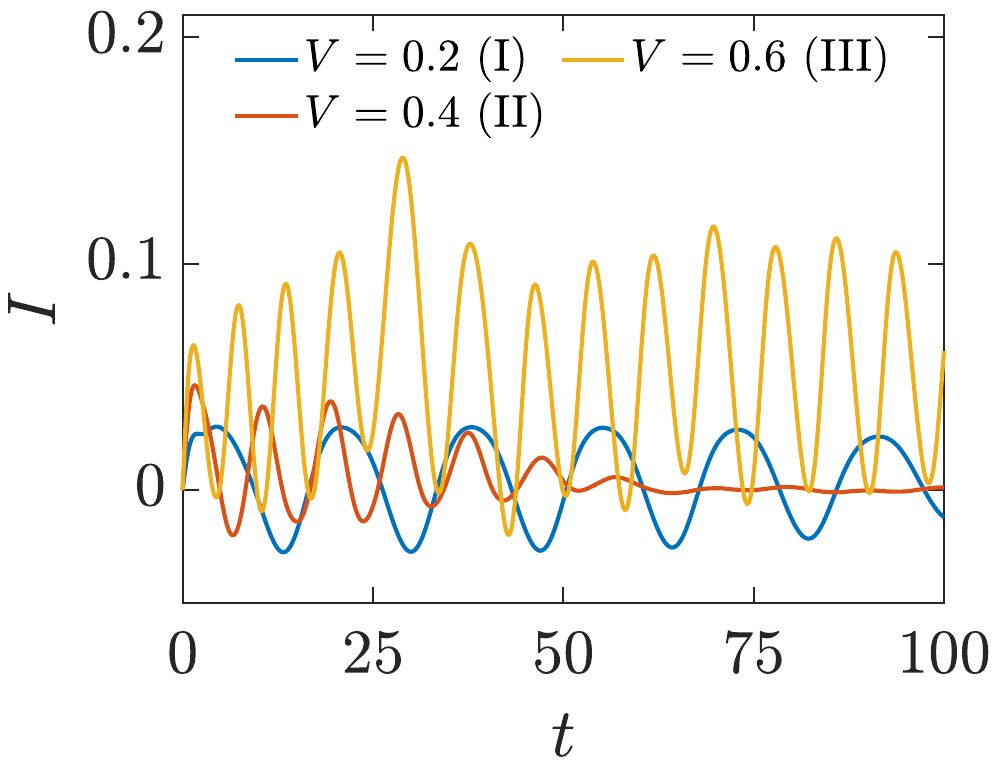}
        \caption{The evolution of the current $I$ in two-lead transport dynamics for various bias voltages $V$, with $J=0.6$, $U=2.0$, and a system size of $N=2L=400$.}
        \label{cur_2to3}
\end{figure}

\subsection{Crossing from regime I to III [cf.~\cref{fig_mid_cur}(b)] by increasing the bias voltage $V$}
\label{apd_cV_23}

As demonstrated in \cref{trans_db}, charge transport in the doublet phase exhibits three distinct regimes, governed by variations in the Kondo coupling $J$. In regime I, the AC Josephson effect is observed. In regime II [cf.~\cref{ld2_reg2}], charge density fluctuations induce a dynamical transition from SC to CDW order, resulting in a damped AC Josephson current, potentially followed by a transient current peak due to enhanced impurity-bath spin correlation. In regime III [cf.~\cref{ld2_reg3}], the current exhibits stable AC and DC components. By increasing $J$, the system transitions from regime I to regime III. Here, we show that fixing $J$ and increasing the voltage $V$ also drives the system from regime I to regime III.

This transition is illustrated in \cref{cur_2to3}. With $J=0.6$ held constant and $V$ increased, the system exhibits an AC Josephson current with slow damping at $V=0.2$ (regime I) and a more rapidly damped AC Josephson current at $V=0.4$ (regime II), as higher voltage accelerates the SC to CDW transition. When $V$ is further increased to $V=0.6$, the DC component becomes sufficient to alter the bath's filling, leading the system into regime III, characterized by a sustained AC Josephson current alongside a DC component. This result, together with that in \cref{fig_mid_cur}(c) (tuning $J$ while fixing $V$), leads to the transport phase diagram in the doublet phase shown in \cref{fig_mid_cur}(b).

\end{document}